\documentclass[a4paper]{amsart}
\usepackage{amssymb}
\usepackage{amsmath}
\usepackage{amsfonts}
\usepackage{hyperref}
\usepackage{graphicx}
\usepackage{subfig}
\usepackage{array}

\def\Eq #1{{\begin{equation} #1 \end{equation}}}
\def\set #1{\{\, #1 \,\}}   
\def\pois #1 #2{\{\,#1,\,#2\,\}}

\def\C{\mathbb{C}}

\def\Z{{\mathbb{Z}}}    

\def\IZ{\mathbb{Z}}

\def\CC{\mathcal{C}}

\def\CO{\mathcal{O}}
\def\CM{\mathcal{M}}

\def\CH{\mathcal{H}}

\def\CU{\mathcal{U}}

\def\CN{\mathcal{N}}

\def\Tr{{\mathrm{Tr \,}}} 
\def\d{{\mathrm{d}}} 
\def\ran{\mathrm{Ran\, }}
\def\vol{{\mathrm{vol\,}}}   
\def\dom{{\mathrm{Dom\, }}}
\def\pr{{\mathrm{pr\,}}}
\def\mitad{{\frac{1}{2}}}   
\def\Cl{{\mathrm{Cl\, }}}

\def\res{{\mathrm{res}}}   
\def\ellip{\mathrm{ellip}}
\def\dim{\hbox{{\rm dim}}}              
\def\Hilbert{\mathcal{H}}  

\newtheorem{theorem}{Theorem}

\newtheorem{proposition}{Proposition}

\newtheorem{lemma}{Lemma}

\begin{document}

\title[Self--adjoint extensions of Laplace--Beltrami and Dirac operators]{Three lectures on global boundary conditions and the theory of self--adjoint extensions of the covariant Laplace--Beltrami and Dirac operators on Riemannian manifolds with boundary \\ {\small{\rm Fall Workshop on Geometry and Physics, ICMAT 2011}} }

\subjclass[2000]{02.30.Tb, 02.40.Vh, 03.65.-w, 03.65.Db}

\keywords{Self--adjoint extensions, Laplace--Beltrami operator, Dirac operator, Self--adjoint Grassmannian}

\author{A. Ibort}\address{Departamento de Matem\'aticas, Universidad Carlos III de Madrid\\ Avda. de la Universidad 30, 28911 Legan\'es, Madrid, Spain}
\email{albertoi@math.uc3m.es}

\begin{abstract}  
In these three lectures we will discuss some fundamental aspects of the theory of self--adjoint extensions of the covariant Laplace--Beltrami and Dirac operators on compact Riemannian manifolds with smooth boundary emphasizing the relation with the theory of global boundary conditions.  

Self--adjoint extensions of symmetric operators, specially of the Laplace--Beltrami and Dirac operators, are fundamental in Quantum Physics as they determine either the energy of quantum systems and/or their unitary evolution.    The well--known von Neumann's theory of self--adjoint extensions of symmetric operators is not always easily applicable to differential operators, while the description of extensions in terms of boundary conditions constitutes a more  natural approach.  Thus an effort is done in offering a description of self--adjoint extensions in terms of global boundary conditions showing how an important family of self--adjoint extensions for the Laplace-Beltrami and Dirac operators are easily describable in this way.  

Moreover boundary conditions play in most cases an significant physical role
and give rise to important physical phenomena like the Casimir effect.  
The geometrical and topological structure of the space of global boundary conditions determining regular self--adjoint extensions for these fundamental differential operators is described.   It is shown that there is a natural homology class dual of the Maslov class of the space.  

A new feature of the theory that is succinctly presented here is the relation between topology change on
the system and the topology of the space of self--adjoint extensions of its Hamiltonian.    Some examples will be commented and the one--dimensional case will be thoroughly discussed.
\end{abstract}
 
\maketitle

\tableofcontents


\section{Introduction}

A simple description of the evolution of a closed autonomous quantum system is provided
by a one--parameter group of unitary operators on a Hilbert space.
Because of Stone's theorem, there is a one--to--one correspondence between 
one--parameter groups of unitary operators on Hilbert spaces and self--adjoint
(not bounded in general) operators.    

Unfortunately as it happens in many occasions, the
operators that we would like to use to describe a quantum system are not
self--adjoint but merely symmetric, hence the evolution defined by them will not
be unitary in general.     The lack of unitarity reflects the fact that there is a ``leak''
of probability because the system is not truly closed but it is, for instance,  in 
interaction with another system.     

The interaction with another external system,
whose detailed form is not known in general, is simulated in many cases by 
introducing a boundary in the system.   It could also happen that the boundary is
introduced just to make sense of the system as it cannot be defined on an infinite domain (for instance when we consider a system on a ``box'').

In all these situations the construction of quantum mechanical systems requires a
detailed analysis of the boundary conditions (BC) imposed on the
system.  Such boundary conditions are fundamental to construct the dynamics of the system and are either determined by the observers and the
experimental setting used by them, or are inherent to the system under consideration. 
This is a common feature of all quantum systems, even the simplest ones. 

The determination of the boundary conditions, or the self--adjoint extensions
of families of symmetric operators, will affect not only the dynamical
evolution of the system, but also the results of the measures realized on the system because
the measurable quantities of the system, which are defined by self--adjoint operators, will depend on them and the spectrum of the quantum observables will vary with the chosen self--adjoint extension.  

Among the most conspicuous quantum systems we find the  ``free'' motion
on a Riemannian manifold.  In such case, the self--adjoint operator defining
the unitary evolution of the system on the Hilbert space of square integrable 
functions on the manifold is the Laplace--Beltrami operator determined by
the Riemannian metric.   However if the manifold has boundary the Laplace--Beltrami operator
is merely symmetric, and to determine completely such quantum system, a 
self--adjoint extension must be chosen.   It is clear that selecting a self--adjoint extension 
in this case must be related to determine the behavior of the system when it ``reaches''
the boundary.     Fixing the behavior of the system at the boundary would then determine its quantum
evolution and in many occasions has a direct physical interpretation. 

Dirac's Correspondence Principle \cite{Di64} provides a useful tool to analyze 
fundamental aspects of quantum mechanical
systems by looking at their classical counterparts (in case they exist), but it is not obvious at all how it extends to include boundary conditions.  
For instance, Dirichlet boundary conditions corresponds to impenetrability of the classical walls
determining the boundary of the classical system in configuration space
but, what are the corresponding classical conditions for mixed BC?   

Conversely, we can address the question of quantizing classical BC.  In
particular we may ask if the classical determination of BC is enough to
fully describe a quantized system.   As the experimental and
observational capabilities are getting more and more powerful, we are
being forced to consider general boundary conditions beyond the
classical cases, Dirichlet, Neumann, etc.   In condensed matter models
``sticky'' boundary conditions have proved to be useful in understanding
the Quantum Hall effect \cite{Jo95}; in quantum
gravity, self--adjoint extensions are used to understand signature change
\cite{Eg95}.  Even at a more fundamental level, topology change in quantum systems
has been modelled using dynamics on BC \cite{Ba95}.   

Following Dirac's approach, we can develop a canonical quantization
program for classical systems with boundary.   Such program requires
a prior discussion on the dynamics of Hamiltonian systems with
boundary.   Without entering a full discussion of this, we may assume that a classical Hamiltonian system with
boundary is specified by a Hamiltonian function $H$ defined on the phase
space $T^*\Omega$ of a configuration space $\Omega$ with boundary
$\partial \Omega$, together with a canonical transformation $S$ of the symplectic
boundary $T^*(\partial \Omega)$ of $T^*\Omega$.  Thus, the classical
boundary conditions (CBC) form a group, the group of symplectic
diffeomorphisms of $T^*(\partial\Omega)$\footnote{Similar
considerations can be made for more general classical phase spaces,
though some care is needed to define their symplectic boundary.}.
Dirac's correspondence principle will be stated now as follows:  Given a
classical BC $S$, and two classical observables $f,g$ on $T^*\Omega$, the corresponding
quantum BC $\hat{S}$ and
self--adjoint operators $\hat{f}_S$, $\hat{g}_S$ depending on
$\hat{S}$, must satisfy:
\begin{equation}\label{DQ1} 
[\hat{f}_S , \hat{g}_S ] = i\hbar \widehat{\pois f g}_S ,
\end{equation} 
and 
\begin{equation}\label{DQ2} 
\hat{S} \circ \hat{R} = \widehat{S R} ,
\end{equation}
where the composition on the left in Eq. \eqref{DQ2} is the group composition on the space of
quantum BC to be discussed later on.  It is obvious that as in the
boundaryless situation, such quantization rules could not be implemented
for all observables and all classical BC.  So, one important question emerging from this analysis is how to select subalgebras of classical observables and
subgroups of classical BC suitable for quantization.  

Before embarking in such
enterprise, some relevant aspects of the classical and the quantum
picture of systems with boundaries have to be clarified.  For instance, we need to understand the
structure of the self--adjoint extensions operators corresponding to a given
classical observable.  The most important class of operators arising in
the first quantization of classical systems are first and second
order elliptic differential operators: for instance, as it was indicated before, the Laplace-Beltrami operator
when quantizing a classical particle without spin, the covariant Laplacian and the Dirac operator
for the quantization of particles with spin.  
This family of operators are
certainly the most fundamental of all elliptic operators (in the Euclidean picture).
Thus for Dirac and Laplace operators we would like to understand their
self--adjoint realizations in terms of classical and  quantum BC.   

Von Neumann developed a general theory of self--adjoint extensions of
symmetric operators in Hilbert spaces \cite{Ne29}.  Such theory is usually presented in the
realm of abstract Hilbert space theory.  This causes that in many cases when applied to discuss differential operators some of the relevant
features attached to the geometry of the operators are
lost.   We will proceed by using a direct approach to describe a large class of self--adjoint extensions of first and second order elliptic
differential operators in terms of global boundary conditions.  Such approach will preserve the geometry of such operators obtaining in this way a fresh interpretation of von
Neumann's theory.  In particular it will be shown that there is a canonical group structure on the space of self--adjoint extensions which is directly related to the boundary conditions imposed on the system as needed for the implementation of Dirac's quantization rule Eq. \eqref{DQ2}.

Elliptic differential operators on compact manifolds have been exhaustively studied
culminating with the celebrated Atiyah--Singer
index theorem that relates the analytical index of such operators with the
topological invariants of the underlying spaces.   Such analysis extends
to manifolds with boundary provided that appropriate elliptic boundary
conditions are used.  A remarkable example is provided by the study of the index of Dirac's operators on manifolds with boundary where appropriate global
elliptic boundary conditions, the so called APS conditions, were introduced in \cite{At71} (see also \cite{Gr95}).
Such extensions have been adequately generalized for higher order
elliptic operators giving rise to interesting constructions of boundary
data \cite{Fr97}.  

The program sketched so far concerns
exclusively first quantization of classical systems, but second
quantization, this is Quantum Field Theory, is needed to truly understand the basic facts of Nature. 
We have already seen that first quantization of classical systems
requires to consider the quantization of boundary conditions, which
leads automatically to consider QBC for the first
quantized system.   Thus even for a very simple system, like a fermion
propagating on a disk we need to consider ``all'' self--adjoint extensions
of the Dirac's operator on the disk.   Thus, to proceed to second
quantization we need to understand the global structure of such space of
extensions.   We will show that such space is a Lagrangian submanifold,
that will be called the self--adjoint Grassmannian,
of the infinite dimensional elliptic Grassmannian manifold.  
Such infinite dimensional Grassmannian was introduced in the study of
integrable hierarchies of nonlinear partial differential equations such
as KdV and KP \cite{Se85}.    It represents a ``universal phase space''
for a large class of integrable evolution problems.  
Our approach here is different, the infinite dimensional
Grassmannian appears as the natural setting to discuss simultaneously
all QBC for a first quantized classical system of arbitrary
dimension.   In fact, a subset of relevant QBC are contained in a submanifold,
the self--adjoint Grassmannian, that should be subjected to second
quantization.  Lagrangian submanifolds of symplectic manifolds play the
role of ``generalized functions'', thus, such programme would imply
quantizing a particular observable of the Grassmannian, making contact in this way
with string theory.  

The theory of boundary value problems for elliptic operators was beautifully described by 
G. Grubb \cite{Gr68}, \cite{Gr03}.  In her work a characterization of all self--adjoint extensions of a class of elliptic operators in terms of global boundary conditions was provided (see also Frey's Ph. D. dissertation where
the boundary value problems for Laplace--Beltrami and Dirac operators was
discussed at length \cite{Fr05}).   However no attempt will be made to relate the results presented in these lectures with Grubb's theory or with the techniques to construct self--adjoint extensions called boundary triples \cite{Br08}.    The reasons for that are two--fold:  on one side we want
to keep the presentation as simple and self--contained as possible and to cover Grubb's theory will be impossible within the scope of these lectures.   On the other hand many of the applications that we were 
referring previously fit perfectly in the framework we are considering here, thus the theory we
are presenting is enough to deal with an important family of problems, even if there are interesting examples that will not be covered by the results in these notes like M. Berry's $D$--singular boundary conditions \cite{Be08}.  

Some of the material presented here has already appeared published elsewhere (see for instance \cite{As05} where some of the preliminary ideas on the global topology of the space of self--adjoint extensions for the covariant Laplacian and its relation to topology change appeared for the first time), or will appear in various forms (see for instance \cite{Ib11} for a detailed discussion of 1D Schr\"odinger operators).   The general theory of self--adjoint extensions from the point of view of quadratic forms is discussed in \cite{Ib12} and will not be considered here as well as the theory of self--adjoint extensions with symmetry that will be discussed elsewhere.


\section{Lecture 1:  Boundary conditions and self--adjoint extensions:  the quest for unitarity}

\subsection{The problem of unitary evolution in Quantum Mechanics}

As it was already pointed it out in the introduction the study of self--adjoint extensions of symmetric operators has its origin in solving the problem of unitary evolution in Quantum Mechanics.
We will discuss briefly some fundamental notions of quantum mechanical systems to 
clarify this point.    

The simplest mathematical formalism to describe an isolated quantum system is by
considering a complex separable Hilbert space $\mathcal{H}$ whose rays are going to be identified with
pure states of the system.   The observables of the system are a family of
self--adjoint operators $\mathcal{O}$ on the Hilbert space $\Hilbert$.     Self--adjoint operators play a double
role in the quantum formalism.  On one side they are observables of the system and its eigenvalues are
the possible outcomes of measures performed on it.   To be more precise, if $A$ is a self--adjoint operator on $\Hilbert$
the spectral theorem states that there exists a projector--valued Borelian spectral measure $E$ on $\mathbb{R}$ such that:
$$ A = \int \lambda\, E(\d\lambda ) .$$

If $|\psi \rangle$ denotes a unitary vector on $\Hilbert$, this is, a representative for a pure state $\rho$ of the system, then the probability
of obtaining an output lying on the Borelian set $\Delta \subset \mathbb{R}$ when measuring the observable $A$ on the state $\rho$
is given by $\mu_{A,\rho}(\Delta) = \int_{\Delta}  \langle \psi | E(\d \lambda)  | \psi \rangle$.  

On the other hand Stone-von Neumann theorem allows to consider self--adjoint operators
 as infinitesimal generators of unitary evolution, i.e. there is a one--to--one correspondence between 
self--adjoint operators $H$ and strongly continuous one--paramenter groups of unitary operators $U_t$ with $H = i \lim_{t\to 0} (U_t - I)/t$ or $U_t = \exp (itH)$.

In both cases, either when we interpret a self--adjoint operator as an observable of the system or when we are constructing the unitary evolution of the system,  the physical interpretation of the operator depends crucially on its self--adjointness.  However in many occasions when 
constructing quantum systems we will need to consider observables or generators of dynamical evolution which are  to be defined by
means of operators that are not self--adjoint but merely symmetric.     Let us recall that if $A$ is a linear operator on $\Hilbert$ which is densely defined with domain $\dom (A)$ then its adjoint operator $A^\dagger$ is uniquely defined and has dense domain $\dom (A^\dagger)$.   A vector $|\psi \rangle$ is in $\dom (A^\dagger)$ if there exists $| \zeta \rangle \in \Hilbert$ such that $\langle \psi, A\phi \rangle = \langle \zeta, \phi \rangle$ for all $|\psi \rangle \in \dom (A)$.
Then the operator $A$ is self--adjoint if $\dom (A) = \dom (A^\dagger)$ and $A = A^\dagger$ on its common domain.   If $\dom (A)$ is merely contained in $\dom (A^\dagger)$ and $A^\dagger \mid_{\dom (A)} = A$, then the operator $A$ is said to be symmetric.   An operator $B$ with domain $\dom (B)$ is said to be an extension of the operator $A$ if $\dom (A) \subset \dom (B)$ and $B \mid_{\dom (A)} = A$.  In this sense, an operator $A$ is said to be symmetric if $A^\dagger$ is a (strict) extension of $A$.

It is easy to provide examples of
symmetric operators for which the spectral theorem fails (in the form above).   In fact, it can be shown easily  that there are symmetric operators whose spectrum is the full field of complex numbers.  Thus we conclude that if we pretend to describe a physical observable or define the unitary evolution of a quantum system, we may not use symmetric operators but self--adjoint ones.   

The following problem arises immediately:  given a symmetric operator $A$ on a Hilbert space $\Hilbert$, does there exists a self--adjoint operator extending it and, if this were the case, how many different self--adjoint extensions do there exists?  

Notice that both parts, the existence and the (non--)uniquenes, of the problem are relevant.  In fact, if
an observable of a quantum system is constructed starting from a symmetric operator $A$, we will not be able to interpret the results of performing measurements of such observable until we have made precise which self--adjoint extension of the symmetric operator $A$ is actually representing the observable we are measuring.  Notice that different self--adjoint extensions of the symmetric operator $A$ have different spectrum, then the expected measurements (this is the physical predictions) would be different.   

Similarly, two different self--adjoint extensions would lead to different unitary evolution groups, thus the prediction of how a given quantum state will evolve would depend on the self--adjoint extension we choose.  Actually it could even happen that a symmetric operator will have no self--adjoint extension at all, then the attempt to describe such observable will be futile.

In this lecture we will describe some aspects regarding the solution of both problems for the (covariant) Laplace--Beltrami operator.    The Laplace--Beltrami operator is used to describe the energy (the Hamiltonian) as well as the infinitesimal generator of unitary evolution of a large class of quantum systems.     

\subsection{The Laplace--Beltrami operator}\label{Laplace_Beltrami}

As it was stated in the introduction we will restrict our attention to the simpler case of Schr\"odinger operators on compact manifolds with smooth boundary and regular potentials.       The Schr\"odinger operator for a particle of mass $m$ moving on a smooth manifold $\Omega$ with boundary $\partial \Omega$ and riemannian metric $\eta$ is given by the Hamiltonian operator $H$ that, in local coordinates $x^i$, takes the form:
\begin{equation}\label{schrodinger}
 H = -\frac{\hbar^2}{2 m} \frac{1}{\sqrt{| \eta |}} \frac{\partial }{\partial x^j} \sqrt{| \eta | }\eta^{jk} \frac{\partial }{\partial x^k} + V(x) ,
 \end{equation}
with the metric tensor $\eta$ given by $\eta = \eta_{jk} (x) dx^k dx^j$, $| \eta | = | \det{ \eta_{jk}(x)} | $ and $\eta^{ij} \eta_{jk} = \delta_k^i$.
The second order differential operator\footnote{From now on we will assume that $\hbar = m = 1$.}
\begin{equation}\label{laplace}
 \Delta_\eta =   \frac{1}{\sqrt{| \eta |}} \frac{\partial }{\partial x^j} \sqrt{| \eta | }\eta^{jk} \frac{\partial }{\partial x^k} 
 \end{equation} 
is formally self--adjoint in the sense that
\begin{equation*}\label{formal}
\langle \Psi, \Delta_\eta \Phi \rangle = \int_\Omega \bar{\Psi}\,  \Delta_\eta \Phi \, \vol_n = - \int_\Omega ( \d\Psi , \d\Phi )_\eta \vol_\eta = \int_\Omega (\overline{\Delta_\eta \Psi}) \, \Phi \, \vol_\eta = \langle \Delta_\eta \Psi, \Phi \rangle  ,
\end{equation*}
for $\Psi$, $\Phi$ any smooth complex valued functions with compact support contained in $\Omega \setminus \partial \Omega$.  In the previous formula $\vol_\eta$ denotes the riemannian volume form on $\Omega$ defined by $\eta$, i.e., $\vol_\eta = \sqrt{|\eta|} dx^1 \wedge \cdots \wedge dx^n$, and $(\d\Psi, \d\Phi)_{\eta(x)} = \eta^{jk}(x) \partial \overline{\Psi}/\partial x^j \partial \Phi/\partial x^k$ is the inner product among covectors at $x\in \Omega$.    In fact, the differential expression (\ref{laplace}) defines a symmetric operator on the space $L^2(\Omega)$ of square integrable functions on $\Omega$ with respect to the measure defined by the volume form $\vol_\eta$,  with dense domain $C_c^\infty (\Omega \setminus \partial \Omega)$ the space of smooth functions with compact support in $\Omega \setminus \partial \Omega$.   The operator $\Delta_\eta$ is closable and its closed extension (the minimal extension such that its graph is closed) has a domain given by 
the closure of $C_c^\infty (\Omega \setminus \partial \Omega)$ with respect to the Sobolev norm $|| \cdot ||_{2,2}$\footnote{The Sobolev norm $|| \cdot ||_{2,2}$ is defined by: $|| \Psi ||_{2,2}^2 = || \Psi ||_{L^2(\Omega)}^2 + || \d\Psi ||_{L^2(\Omega)}^2$ where $|| \d\Psi ||_{L^2(\Omega)}^2 = \int_{\Omega} ( \d\Psi , \d\Psi )_\eta \vol_\eta$.}.  
We will denote this minimal extension of the operator $\Delta_\eta$ by $\Delta_0$ and its domain by $\mathcal{D}_0$.  Notice that $\mathcal{D}_0$ is the space  $\mathcal{H}_0^2(\Omega)$ of functions of Sobolev class 2 on $\Omega$ vanishing at the boundary together with its normal derivative, i.e., the space of functions on $L^2(\Omega)$ possessing first and second weak derivatives that are square integrable and such that their boundary values vanish (see Lions trace theorem below Thm. \ref{trace}).  

 The operator $\Delta_0$ is symmetric but not self--adjoint on $L^2(\Omega)$ because the adjoint operator  $\Delta_0^\dagger$ has a dense domain $\dom (\Delta_0^\dagger) =: \mathcal{D}_0^\dagger$ that contains strictly $\mathcal{D}_0$.
The domain $\mathcal{D}_0^\dagger$ contains the space $\mathcal{H}^2(\Omega)$ of all functions of Sobolev class 2 on $\Omega$, i.e., the space of functions on $L^2(\Omega)$ possessing first and second weak derivatives that are square integrable.   Self--adjoint extensions of the operator $\Delta_\eta$ are given by operators $\Delta_\mathcal{D}$ with domain $\mathcal{D}$ such that $\mathcal{D}_0 \subset \mathcal{D} \subset \mathcal{D}_0^\dagger$, $\Delta_\mathcal{D}\mid _{\mathcal{D}_0} = \Delta_0$ and $\Delta_\mathcal{D} = \Delta_\mathcal{D}^\dagger$.  Notice that in that case $\Delta_\mathcal{D} = \Delta_0^\dagger\mid_\mathcal{D}$.    

Von Neumann's theorem establishes (see for instance \cite{We80} Thm. 8.12, and Lecture 2, Thm. \ref{von_Neumann} these notes) that there is a one--to--one correspondence between self--adjoint extensions $\Delta_\mathcal{D}$ of the Laplace--Beltrami operator $\Delta_0$ and unitary operators $K \colon \mathcal{N}_+ \to \mathcal{N}_-$, where the deficiency spaces $\mathcal{N}_\pm$ are defined as:
$$ \mathcal{N}_\pm = \{ \Psi \in L^2(\Omega )  \mid  \Delta_0^\dagger \Psi = \pm i \Psi \} .$$
In particular, given the unitary operator $K$, the domain $\mathcal{D}$ of the operator $\Delta_\mathcal{D}$ is given by $\mathcal{D} = \mathcal{D}_0 \oplus (I + K)\mathcal{N}_+$, and the extended operator $\Delta_\mathcal{D}$ takes the explicit form:
$$ \Delta_\mathcal{D} (\Psi_0 \oplus (I+ K)\xi_+ ) = \Delta_0\Psi_0 \oplus i(I-K) \xi_+ ,$$
for all $\Psi_0 \in \mathcal{D}_0$ and $\xi_+\in \mathcal{N}_+$.
 
\subsection{Self--adjoint extensions of the Laplace--Beltrami operator and boundary conditions}

Unfortunately, as it was stated in the introduction, von Neumann's theorem is not always well suited for the explicit construction of general self--adjoint extensions of the Laplace--Beltrami operator (it is necessary to  determine first the deficiency spaces $\mathcal{N}_\pm$ that could be difficult).  We can take however a different route inspired in the classical treatment of formally self--adjoint differential operators.   If we rewrite the identity expressing the formal self--adjointness of $\Delta_\eta$ for functions $\Psi$, $\Phi$ in $C^\infty(\overline{\Omega})$ instead of $C_c^\infty(\Omega \setminus \partial \Omega)$, a simple computation shows:

\begin{equation}\label{boundary}
 \int_\Omega \bar{\Psi} \, \Delta_\eta \Phi \, \vol_\eta = \int_\Omega (\overline{\Delta_\eta \Psi})\,  \Phi \, \vol_\eta + \int_{\partial \Omega} \left( \bar{\psi} \dot{\varphi} - \dot{\bar{\psi}} \varphi \right) \vol_{\partial \eta},
\end{equation}
where $\psi = \Psi\mid_{\partial \Omega}$, $\varphi = \Phi\mid_{\partial \Omega}$, and the normal derivative $\dot{\varphi}$ is defined
as:
$$ \dot{\varphi} \, \vol_{\partial \eta} =   \star \mathrm{d}\,\Phi\mid_{\partial \Omega} $$
where $\star$ is the Hodge operator defined by the metric $\eta_{ij}$ and $\vol_{\partial \eta}$ is the riemannian volume defined on the boundary $\partial \Omega$ by the restriction $\partial \eta$ of the Riemannian metric $\eta$ to it.   Less intrinsically, but more explicitly, we have $\dot{\varphi} = \frac{d \Phi}{d \nu}\mid_{\partial \Omega} = \d\Phi (\nu )$ where $\nu$ is the exterior normal vector to $\partial \Omega$.  We obtain the Lagrange boundary form $\Sigma$ for the Laplace--Beltrami operator:
\begin{equation}\label{lagrange}
\Sigma ((\psi,\dot{\psi}), (\varphi,\dot{\varphi})) = \langle \psi, \dot\varphi \rangle_{L^2(\partial \Omega)} - \langle \dot{\psi}, \varphi \rangle_{L^2(\partial \Omega)} .
 \end{equation}
In what follows, if there is no risk of confusion, we will omit the subscript $L^2(\partial \Omega)$ that denotes the $L^2$ inner product on the boundary manifold $\Gamma = \partial \Omega$ with respect to the measure defined by the volume form $\vol_{\partial \eta}$, hence we will simply write $\langle \psi, \varphi \rangle = \int_{\partial \Omega} \bar{\psi} \varphi \,  \vol_{\partial \eta}$.    The Lagrange boundary bilinear form $\Sigma$ defines a continuous bilinear form on the Hilbert space $L^2(\partial \Omega)\oplus L^2(\partial \Omega)$:
\begin{equation}\label{lagrange2}
\Sigma ((\psi_1,\psi_2), (\varphi_1,\varphi_2)) = \langle \psi_1, \varphi_2 \rangle - \langle \psi_2, \varphi_1 \rangle,  \end{equation}
for all $(\psi_1,\psi_2), (\varphi_1,\varphi_2) \in  L^2(\partial \Omega)\oplus L^2(\partial \Omega)$.

If we denote by $\gamma \colon C^\infty (\Omega) \to C^\infty(\partial \Omega)\oplus C^\infty(\partial \Omega)$ the trace map given by $\gamma (\Psi ) = (\psi, \dot{\psi})$, the Lions-Mag\'enes trace theorem \cite{Li72} shows that there exists a continuous extension of $\gamma$ to $\mathcal{H}^2(\Omega)$, actually we have\footnote{It is worth to check for instance Adams' proof in \cite{Ad75}, Thms. 7.50-55.}:

\begin{theorem}\label{trace}
There exists a unique extension of the trace map $\gamma$ to a continuous surjective linear map (denoted with the same symbol)  $\gamma \colon \mathcal{H}^2(\Omega) \to \mathcal{H}^{3/2}(\partial \Omega) \oplus \mathcal{H}^{1/2}(\Omega)$.  Moreover the induced map $\tilde{\gamma} \colon \mathcal{H}^2(\Omega)/\ker \gamma \to \mathcal{H}^{3/2}(\partial \Omega) \oplus \mathcal{H}^{1/2}(\Omega)$ is a homeomorphism and $\ker \gamma = \mathcal{H}_0^2(\Omega)$.
\end{theorem}

We will denote also, as it is customary, the fractional power Sobolev space $\mathcal{H}^{3/2}(\partial \Omega)$ by $W^{3/2,2}(\partial \Omega)$, etc.  Again, sometimes we will prefer to use the notation $b$ (as ``boundary'' map) for the linear map $\gamma$ (the ``trace'' map) defined in the theorem above.

The previous observations provide a simple characterization of a large class of self--adjoint extensions of the operator $\Delta_0$.   In fact it is easy to check that:

\begin{theorem}\label{self1}
There is a one--to--one correspondence between self--adjoint extensions $\Delta_\mathcal{D}$ of the Laplace--Beltrami operator $\Delta_0$ with domain $\mathcal{D}$ contained in $\mathcal{H}^2(\Omega)$ and non--trivial maximal closed isotropic subspaces $W$ of the Lagrange form $\, \Sigma$ contained in $W^{3/2,2}(\partial\Omega) \oplus W^{1/2,2}(\partial \Omega)$.    The correspondence being explicitly given by $\mathcal{D} \mapsto W = \gamma (\mathcal{D})$.
\end{theorem}

{\parindent 0cm \emph{Proof:}}  
Let $\mathcal{D}\subset \mathcal{H}^2(\Omega)$ be the domain of a self--adjoint extension $\Delta_\mathcal{D}$ of the operator $\Delta_0$.  Consider the subspace $W := \gamma (\mathcal{D})  \subset W^{3/2,2}(\partial \Omega) \oplus W^{1/2,2}(\partial \Omega) \subset  L^2(\partial \Omega)\oplus L^2(\partial \Omega)$ consisting on the set of pairs of functions $(\varphi, \dot{\varphi})$ that are respectively the restriction to $\partial \Omega$ of a function $\Phi\in \mathcal{D}$ and its normal derivative.  Notice that the subspace $\mathcal{D}$  is closed in $\mathcal{H}^2(\Omega)$.  Hence, because $\gamma$ is an homeomorphism from $\mathcal{H}^2(\Omega)/\ker \gamma$ to $W^{3/2,2}(\partial \Omega) \oplus W^{1/2,2}(\partial \Omega) $, $\gamma (\mathcal{D})$ is a closed subspace of $W^{3/2,2}(\partial \Omega) \oplus W^{1/2,2}(\partial \Omega)$.   Because of Eq. \eqref{boundary}, it is clear that $\Sigma\mid_{W} = 0$ and the subspace $W$ is maximally isotropic in $W^{3/2,2}(\partial \Omega) \oplus W^{1/2,2}(\partial \Omega)$, because if this were not the case, there will be a closed  isotropic subspace $W' \subset W^{3/2,2}(\partial \Omega) \oplus W^{1/2,2}(\partial \Omega)$  containing $W$.   Then $\gamma^{-1}(W')$ will define a domain $\mathcal{D}'$ containing $\mathcal{D}$ such that  the operator $\Delta_0$ would be symmetric on it, in contradiction with the self--adjointness assumption.

Conversely, let $W\subset L^2(\partial \Omega)\oplus L^2(\partial \Omega)$ be a maximal closed $\Sigma$--isotropic subspace in $W^{3/2,2}(\partial \Omega) \oplus W^{1/2,2}(\partial \Omega)$.  Then consider the closed subspace $\mathcal{D}_W : =  \gamma^{-1}(W) \subset \mathcal{H}^2(\Omega)$ of functions $\Psi$ such that $(\psi, \dot{\psi}) \in W$.   It is clear that for any pair of functions $\Psi$, $\Phi$ on $\mathcal{D}_W$, because $W$ is isotropic with respect to $\Sigma$, then Eq. \eqref{boundary} gives $\langle \Psi, \Delta_\eta \Phi \rangle = \langle \Delta_\eta \Psi,\Phi \rangle$ and the operator $\Delta_\eta$ is symmetric in $D_W$.  Moreover, because of the maximality of $W$in $W^{3/2,2}(\partial \Omega) \oplus W^{1/2,2}(\partial \Omega)$  it is easy to see that $\mathcal{D}_W = \mathcal{D}_W^\dagger$, hence it is self--adjoint.   
\hfill$\Box$
 
\bigskip
 
We could object that the previous characterization of self--adjoint extensions of the Laplace--Beltrami operator in terms of closed maximal isotropic subspaces of $\Sigma$ in $W^{3/2,2}(\partial \Omega) \oplus W^{1/2,2}(\partial \Omega)$ is rather obscure.   An important observation in this sense is that the linear transformation $C \colon  L^2(\partial \Omega)\oplus L^2(\partial \Omega) \to L^2(\partial \Omega)\oplus L^2(\partial \Omega)$, defined by: 
\begin{equation}\label{first_cayley}
C(\varphi, \dot{\varphi}) = \left( \frac{1}{\sqrt{2}}(\varphi+i\dot{\varphi}), \frac{1}{\sqrt{2}}(\varphi-i\dot{\varphi})\right),
\end{equation}
transforms maximally isotropic closed subspaces of $\Sigma$ into graphs of unitary operators of $L^2(\partial \Omega)$.  

\begin{theorem} \label{self2} 
The Cayley map $C$ provides a one--to--one correspondence between maximally isotropic closed subspaces of the Lagrange bilinear boundary form $\Sigma$ in $L^2(\partial \Omega)\oplus L^2(\partial \Omega)$ and graphs  $G_V = \{ (\psi_+, V\psi_+) \mid \psi_+ \in L^2(\partial M) \} $ of unitary operators $V\colon L^2(\partial \Omega) \to L^2(\partial \Omega)$.
\end{theorem}

{\parindent 0cm \emph{Proof:}}    
Notice first that the map $C$ is a unitary operator on the Hilbert space $L^2(\partial \Omega)\oplus L^2(\partial \Omega)$ with the inner product $\langle \langle (\psi_1, \psi_2), (\varphi_1, \varphi_2) \rangle\rangle = \langle \psi_1,\varphi_1 \rangle + \langle \psi_2,\varphi_2 \rangle$.   

Consider now the transformed bilinear form $\tilde{\Sigma}$ on $L^2(\partial \Omega)\oplus L^2(\partial \Omega)$ defined by 
$$\tilde{\Sigma}(C(\varphi,\dot{\varphi}),C(\psi,\dot{\psi})) = \Sigma(( \varphi,\dot{\varphi}),(\psi,\dot{\psi})).$$ 
Thus, using the notation $\varphi_\pm = \frac{1}{\sqrt{2}}(\varphi\pm i \dot{\varphi})$, we have:
$$ \tilde{\Sigma}((\varphi_+,\varphi_-),(\psi_+,\psi_-))= -i \left[  \langle \varphi_+,\psi_+\rangle - \langle \varphi_-,\psi_-\rangle \right] .$$
Hence, if $W$ is a maximally isotropic closed subspace for $\Sigma$, then $\tilde{W}=C(W)$ will be a maximally isotropic closed subspace for $\tilde{\Sigma}$.     Then it is easy to show that $\tilde{W}$ defines the graph of a linear operator.   We first realize that  $(\{ 0 \} \times L^2) \cap \tilde{W} = 0$ because if $(0,\psi_-)$ lays in $\tilde{W}$, then $\tilde{\Sigma}((0,\psi_-),(0,\psi_-)) = i || \psi_- ||^2$ must vanish.   Hence if $(\psi_+,\psi_-)$, $(\psi_+,\hat{\psi}_-)$ are in $\tilde{W}$, then $(0, \psi_--\hat{\psi}_-)$ is in $\tilde{W}$ and $\psi_- = \hat{\psi}_-$.  Then we define a linear operator $V\colon \mathcal{V} \to L^2$ by $U(\psi_+) = \psi_-$ with $(\psi_+,\psi_-) \in \tilde{W}$ and $\mathcal{V}$ the closed subspace of vectors $\psi_+$ such that there exists $(\psi_+,\psi_-)\in \tilde{W}$.  Similarly we can construct another operator $\tilde{V} \colon \tilde{\mathcal{V}} \to L^2$ by observing that $(L^2 \times \{ 0 \} )\cap \tilde{W} = 0$. Then it is easy to show that $V$ is an isometry from $\mathcal{V}$ to $\tilde{\mathcal{V}}$, $\tilde{V}$ is an isometry from $\tilde{\mathcal{V}}$ to $\mathcal{V}$ and they are inverse of each other.   Then because of the maximality of $\tilde{W}$, we conclude that $\mathcal{V} = \tilde{\mathcal{V}} = L^2$.
\hfill$\Box$
 
 \bigskip
 
Hence a convenient way of constructing self--adjoint extensions of the Laplace operator will be provided by unitary operators $U$ on $L^2(\partial \Omega)$ such that the preimage under $C$ of their graphs will be closed in $W^{3/2,2}(\partial \Omega) \oplus W^{1/2,2}(\partial \Omega)$.    We will develop this programme in the 1D case in the forthcoming section. 
 
We will end this discussion by realizing that the operator multiplication by a regular function is essentially self--adjoint and its unique self--adjoint extension has domain $L^2(\Omega)$.  Hence, the self--adjoint extensions of the Schr\"odinger operator $H$, Eq. \eqref{schrodinger}, coincide with the self--adjoint extensions of $\Delta_\eta$.    

We can summarize the preceding analysis by stating that under the conditions above, the domain $\mathcal{D}$ of a self--adjoint extension of the Schr\"odinger operator $H$ defined by a closed subspace of functions $\Psi$ on $\mathcal{H}^2(\Omega)$ must satisfy:
\begin{equation}\label{asorey}
\varphi - i\dot{\varphi} = U (\varphi + i \dot{\varphi})
\end{equation}
for a given unitary operator $U \colon L^2 (\partial \Omega) \to L^2(\partial \Omega)$.  The formula above, Eq. \eqref{asorey}, provides a powerful and effective computational tool to deal with large family of self--adjoint extensions of Schr\"odinger operators. It was introduced in a slightly different context by Asorey {\it et al} \cite{As05} and will be used extensively in the rest of this paper.
In what follows we will denote respectively by $H_U$ or $H_\mathcal{D}$ the self--adjoint extension determined by the unitary operator $U$ or the self--adjoint extension whose domain is $\mathcal{D}$. Also according to Thms. \ref{self1} and \ref{self2} we denote such domain as $\mathcal{D}_U$. 
Notice that $U = \mathbb{I}$ corresponds to Neumann's boundary conditions and $U = - \mathbb{I}$ determines Dirichlet's boundary conditions.

 
\subsection{The unitary group of self--adjoint extensions in 1D}

\subsubsection{Self--adjoint extensions of Schr\"odinger operators in 1D}
We will concentrate our attention in 1D were we will be able to provide an elegant formula to solve the spectral problem for each self--adjoint extension.

Notice first that a compact 1D manifold $\Omega$ consists on a finite number of closed intervals $I_\alpha$, $\alpha = 1,\ldots,n$.  Each interval will have the form $I_\alpha = [a_\alpha, b_\alpha] \subset \mathbb{R}$ and the boundary of the manifold $\Omega = \sqcup_{\alpha=1}^n [a_\alpha, b_\alpha]$ (disjoint union) is given by the family of points $\{ a_1, b_1, \ldots, a_n,b_n\}$.     Functions $\Psi$ on $\Omega$ are determined by vectors $(\Psi_1, \ldots, \Psi_n)$ of complex valued functions $\Psi_\alpha \colon I_\alpha \to \mathbb{C}$.

 A Riemannian metric $\eta$ on $\Omega$ is given by specifying a Riemannian metric $\eta_\alpha$ on each interval $I_\alpha$, this is, by a positive smooth function $\eta_\alpha(x) > 0$ on the interval $I_\alpha$, i.e., $\eta\mid_{I_\alpha} = \eta_\alpha (x) dx^2$.    Then the inner product on $I_\alpha$ takes the form $\langle \Psi_\alpha , \Phi_\alpha \rangle = \int_{a_\alpha}^{b_\alpha} (x) \bar{\Psi}_\alpha (x) \Phi_\alpha (x) \sqrt{\eta_\alpha (x)} dx$ and the Hilbert space of square integrable functions on $\Omega$ is given by $L^2(\Omega) = \bigoplus_{\alpha= 1}^n L^2(I_\alpha, \eta_\alpha)$.    Thus the Hilbert space $L^2(\partial \Omega)$ at the boundary reduces to $\mathbb{C}^{2n}$, as well as the subspaces $W^{3/2,2}(\partial \Omega)$ and $W^{1/2,1}(
\partial \Omega)$.  The vectors in $L^2(\partial \Omega)$ are determined by the values of $\Psi$ at the points $a_\alpha$, $b_\alpha$ (with the standard inner product):
 $$\psi = (\Psi_1(a_1),\Psi_1(b_1), \ldots, \Psi_n(a_n),\Psi_n(b_n)).$$  
 Similarly we will denote by $\dot{\psi}$ the vector containing the normal derivatives of $\Psi$ at the boundary, this is:
 $$ \dot{\psi} = \left(  - \left. \frac{d\Psi_1}{dx}\right|_{a_1} , \left. \frac{d\Psi_1}{dx}\right|_{b_1}, \ldots, - \left.\frac{d\Psi_n}{dx}\right|_{a_n},\left.\frac{d\Psi_n}{dx}\right|_{b_n}  \right) .$$
 
Because of Thms. \ref{self1} and \ref{self2}, an arbitrary self--adjoint extension of the Schr\"odinger operator $H$ defined by the riemannian metric $\eta$ and a regular potential function $V$ is defined by a unitary operator $V \colon \mathbb{C}^{2n} \to \mathbb{C}^{2n}$.  Its domain consists of those functions whose boundary values $\psi$, $\dot{\psi}$ satisfy Asorey's condition, Eq. (\ref{asorey}).  This equation becomes a finite dimensional linear system for the components of the vectors $\psi$ and $\dot{\psi}$.    Hence the space of self--adjoint extensions is in one--to--one correspondence with the unitary group $U(2n)$ and has dimension $4n^2$.
 
 It will be convenient for further purposes to organize the boundary data vectors $\psi$ and $\dot{\psi}$ in a different way.  Thus, we denote by $\psi_l\in \mathbb{C}^n$ (respec. $\psi_r$) the column vector whose components $\psi_l(\alpha)$, $\alpha = 1, \ldots, n$, are  the values of $\Psi$ at the left endpoints $a_\alpha$, this is $\psi_l(\alpha) = \Psi_\alpha(a_\alpha )$ (respec. $\psi_r(\alpha ) = \Psi_\alpha(b_\alpha )$ are the values of $\Psi$ at the right endpoints).   Similarly we will denote by $\dot{\psi}_l(\alpha ) =   - \frac{d\Psi_\alpha}{dx}\mid_{a_\alpha}$ and $\dot{\psi}_r(\alpha ) =   \frac{d\Psi_\alpha}{dx}\mid_{b_\alpha}$, $\alpha = 1, \ldots, n$.   Hence, the domain of the self--adjoint extension defined by the unitary matrix $U$ will be written accordingly as:
 \begin{eqnarray}\label{asorey_1D}
\psi_l - i\dot{\psi}_l &=& U^{11}(\psi_l + i\dot{\psi}_l) + U^{12}(\psi_r + i\dot{\psi}_r) \\ 
 \psi_r - i\dot{\psi}_r &=& U^{21}(\psi_l + i\dot{\psi}_l) + U^{22}(\psi_r + i\dot{\psi}_r)  \nonumber
 \end{eqnarray}
and $U$ has the block structure:
\begin{equation}\label{block}
 U = \left[  \begin{array}{c|c} U^{11} & U^{12} \\ \hline U^{21} & U^{22}  \end{array}    \right] .
 \end{equation}
 Notice that the unitary matrix $U$ is related to the unitary matrix $V$ above by a permutation, but we will not need its explicit expression here.
 
 Thus in what follows we will use the notation for the boundary data:
 $$ \psi = \left[ \begin{array}{c} \psi_l \\ \psi_r \end{array}\right]; \quad  \dot{\psi} = \left[ \begin{array}{c} \dot{\psi}_l \\ \dot{\psi}_r \end{array}\right]$$
 and Asorey's condition reads again:
 \begin{equation}\label{asorey2}
 \psi - i \dot{\psi} = U (\psi + i \dot{\psi}), \quad \quad  U \in U(2n) .
 \end{equation} 


\subsubsection{The spectral function}
Once we have determined a self--adjoint extension $H_U$ of the Schr\"odinger operator $H$, we can determine the unitary evolution of the system by computing the flow $U_t = \exp(-itH_U/\hbar )$.   It is well--known that the Dirichlet extension of the Laplace--Beltrami operator has a pure discrete spectrum because of the compactness of the manifold and the ellipticity of the operator, hence all self--adjoint extensions have a pure discrete spectrum (see \cite{We80}, Thm. 8.18).  Then the spectral theorem for the self--adjoint operator $H_U$ states:
$$ H_U = \sum_{k = 1}^\infty \lambda_k P_k ,$$
where $P_k$ is the orthogonal projector onto the finite--dimensional eigenvector space $V_k$ corresponding to the eigenvalue $\lambda_k$.  The unitary flow $U_t$ is given by:
$$ U_t = \sum_{k = 1}^\infty e^{-it\lambda_k / \hbar} P_k .$$
Hence all that remains to be done is to solve the eigenvalue problem:
\begin{equation}\label{eigenvalue_problem}
H_U \Psi = \lambda \Psi ,
\end{equation}
for the Schr\"odinger operator $H_U$.     We devote the rest of this section to provide an explicit formula to solve Eq. (\ref{eigenvalue_problem}).   

On each subinterval $I_\alpha = [a_\alpha, b_\alpha]$ the differential operator $H_\alpha = H|_{I_\alpha}$ takes the form 
of a Sturm--Liouville operator 
$$H_\alpha = - \frac{1}{W_\alpha} \frac{d}{dx} p_\alpha(x) \frac{d}{dx} + V_\alpha(x),$$ 
with smooth coefficients $W_\alpha = \frac{1}{2\sqrt{\eta_\alpha}} > 0$ (now and in what follows we are taking the physical constants $\hbar$ and $m$ equal to 1), $p_\alpha(x) = \frac{1}{\sqrt{\eta_\alpha}}$, hence the second order differential equation 
\begin{equation}\label{eigen_alpha}
 H_\alpha \Psi_\alpha = \lambda \Psi_{\alpha} 
 \end{equation}
has a two-dimensional linear space of solutions for each $\lambda$.   We shall denote a basis of solutions of such space as $\Psi_\alpha^\sigma$, $\sigma = 1,2$.  Notice that $\Psi_\alpha^\sigma$ depends differentially on $\lambda$.    Hence
a generic solution of Eq. (\ref{eigen_alpha}) takes the form:
$$ \Psi_\alpha = A_{\alpha,1} \Psi_{\alpha}^1 + A_{\alpha,2} \Psi_{\alpha}^2 .$$
Now it is clear that 
$$\psi_l(\alpha) = \Psi_\alpha (a_\alpha) = A_{\alpha,1} \psi_a^1 (\alpha) + A_{\alpha,2} \psi_a^2(\alpha) .$$
Hence:
$$ \psi_l = A_1 \circ \psi_a^1 + A_2 \circ \psi_a^2 ,$$
where $A_\sigma$, $\sigma = 1,2$, denotes the column vector
$$ A_\sigma = \left[  \begin{array}{c}  A_{1,\sigma} \\ \vdots \\ A_{n,\sigma} \end{array} \right] $$
and $\circ$ denotes the Hadamard product of two vectors, i.e., $(X\circ Y)_\alpha = X_\alpha Y_\alpha$ where $X,Y \in \mathbb{C}^n$.
We obtain similar expressions for $\psi_r$, $\dot{\psi}_l$ and $\dot{\psi}_r$.
With this notation Eqs. (\ref{asorey_1D}) become:
\begin{eqnarray}\label{asorey_eigen}
(\psi_l^1 - i\dot{\psi}_l^1)\circ A_1 + (\psi_l^2 - i\dot{\psi}_l^2)\circ A_2 &=& U^{11}(\psi_l^1 + i\dot{\psi}_l^1)\circ A_1 + U^{11}(\psi_l^2 + i\dot{\psi}_l^2)\circ A_2  \nonumber \\  +  U^{12}(\psi_r^1 + i\dot{\psi}_r^1)\circ A_1  &+& U^{12}(\psi_r^2 + i\dot{\psi}_r^2)\circ A_2
\end{eqnarray}
\begin{eqnarray*}
(\psi_r^1 - i\dot{\psi}_r^1)\circ A_1 + (\psi_r^2 - i\dot{\psi}_r^2)\circ A_2 &=& U^{21}(\psi_l^1 + i\dot{\psi}_l^1)\circ A_1 + U^{21}(\psi_l^2 + i\dot{\psi}_l^2)\circ A_2  \\ \nonumber +  U^{22}(\psi_r^1 + i\dot{\psi}_r^1)\circ A_1  &+& U^{22}(\psi_r^2 + i\dot{\psi}_r^2)\circ A_2
\end{eqnarray*}
It will be convenient to use the compact notation $\psi_{l\pm}^\sigma = \psi_{l}^\sigma \pm i \dot{\psi}_l^\sigma$, $\sigma = 1,2$, and similarly for  $\psi_{r\pm}^\sigma$.  

If $T$ is a $n\times n$ matrix and $X,Y$ arbitrary $n \times 1$ vectors, we will define $T \circ X$ as the unique matrix such that $(T\circ X)Y = T(X\circ Y)$.    The rows of the matrix $T\circ X$ are $T_i \circ X$ or alternatively, the columns of $T\circ X$ are given by $T^j X_j$ (no summation on $j$).   It can be proved easily that
\begin{equation}\label{hadamardT}
 T\circ X = T\circ (X \otimes \mathbf{1}) ,
 \end{equation}
where $\mathbf{1}$ is the vector whose components are all ones (i.e., the identity with respect to the Hadamard product $\circ$) and the Hadamard product of matrices in the r.h.s. of Eq. \eqref{hadamardT} is the trivial componentwise product of matrices. 
Using these results Eqs. (\ref{asorey_eigen}) become:
\begin{eqnarray*}
(I_n\circ \psi_{l-}^1 - U^{11}\circ \psi_{l+}^1 - U^{12}\circ \psi_{r+}^1  )A_1 + (I_n\circ \psi_{r-}^2 - U^{11}\circ \psi_{l+}^2 - U^{12}\circ \psi_{r+}^2  )A_2 &=& 0 \\
\nonumber  (I_n\circ \psi_{r-}^1 - U^{21}\circ \psi_{l+}^1 - U^{22}\circ \psi_{r+}^1  )A_1 + (I_n\circ \psi_{r-}^2 - U^{21}\circ \psi_{l+}^2 - U^{22}\circ \psi_{r+}^2  )A_2 &=& 0
 \end{eqnarray*}
Thus the previous equations define a linear system for the $2n$ unknowns $A_1$ and $A_2$.  They will have a non trivial solution if and only if the determinant of the $2n \times 2n$ matrix of coefficients $M(U, \lambda)$ below vanish:

$$ M(U, \lambda) = \left[  \begin{array}{cc}  I_n\circ \psi_{l-}^1 - U^{11}\circ \psi_{l+}^1 - U^{12}\circ \psi_{r+}^1 & I_n\circ \psi_{l-}^2 - U^{11}\circ \psi_{l+}^2 - U^{12}\circ \psi_{r+}^2  \\
I_n\circ \psi_{r-}^1 - U^{21}\circ \psi_{l+}^1 - U^{22}\circ \psi_{r+}^1  & I_n\circ \psi_{r-}^2 - U^{21}\circ \psi_{l+}^2 - U^{22}\circ \psi_{r+}^2  \end{array} \right] . $$

The fundamental matrix $M(U, \lambda)$ can be written in a more inspiring form using another operation naturally induced by the Hadamard and the usual product of matrices.    Thus, consider the $2n \times 2n$ matrix $U$ with the block structure of Eq. (\ref{block}) and the $2n \times 2$ matrices:
$$ [\psi_\pm^1 \mid \psi_\pm^2 ]  = \left[   \begin{array}{c|c} \psi_{l\pm}^1 & \psi_{l\pm}^2 \\  \hline \psi_{r\pm}^1 & \psi_{r\pm}^2 \end{array} \right] ,$$
then we define
$$ \left[  \begin{array}{c|c} U^{11} & U^{12} \\ \hline U^{21} & U^{22}  \end{array} \right]  \odot \left[   \begin{array}{c|c} \psi_{l\pm}^1 & \psi_{l\pm}^2 \\ \hline \psi_{r\pm}^1 & \psi_{r\pm}^2 \end{array} \right]  \equiv \left[   \begin{array}{c|c} U^{11}\circ \psi_{l\pm}^1 + U^{12} \circ \psi_{r\pm}^1& U^{11}\circ \psi_{l\pm}^2 + U^{12}\circ  \psi_{r\pm}^2 \\  \hline U^{21}\circ \psi_{l\pm}^1 + U^{22}\circ \psi_{r\pm}^1 & U^{21}\circ\psi_{l\pm}^2 + U^{22} \circ \psi_{r\pm}^2\end{array} \right] 
$$
and similarly
$$ I_{2n} \odot  [\psi_\pm^1 \mid \psi_\pm^2 ]  = \left[  \begin{array}{c|c}  I_n\circ \psi_{l\pm}^1 & I_n \circ \psi_{l\pm}^2 \\
\hline I_n\circ \psi_{r\pm}^1 & I_n \circ \psi_{r\pm}^2   \end{array}\right] .$$
Finally we conclude that the condition for the existence of coefficients $A_1$ and $A_2$ such that the solutions to the eigenvalue equation lie in the domain of the self--adjoint extension defined by $U$ is given by the vanishing of the spectral function $\Lambda_U (\lambda) = \det M(U, \lambda )$, that written with the notation introduced so far becomes:
\begin{equation}\label{spectral_function}
\Lambda_U(\lambda) =  \det ( I_{2n} \odot [\psi_-^1 \mid \psi_-^2 ]  - U \odot  [\psi_+^1 \mid \psi_+^2 ]  ) = 0 .
 \end{equation}
The zeros of the spectral function $\Lambda$ provide the eigenvalues $\lambda$ of the spectral problem Eq. (\ref{eigenvalue_problem}).

In the particular case $n =1$, the previous equation becomes greatly simplified, the Hadamard product becomes the usual scalar product and the Hadamard--matrix product is the usual product of matrices.  After some simple manipulations, the spectral function $\Lambda_U (\lambda)$ becomes:  
\begin{eqnarray}\label{spectral_1}
\Lambda_U (\lambda) &=& W(l,r,-,-) +U^{11}W(r,l,-,+)+U^{22}W(r,l,+,-) \\ &+& U^{12}W(r,r,-,+)  +U^{21}W(l,l,+,-)+\det U\cdot W(l,r,+,+)\notag
\end{eqnarray}
where we have used the notation:
$$ W(l,l,+,-) = \left|  \begin{array}{cc}  \psi_{l+}^1 & \psi_{l+}^2 \\ \psi_{l-}^1 & \psi_{l-}^2  \end{array} \right|, \quad W(l,r,+,-) = \left|  \begin{array}{cc}  \psi_{l+}^1 & \psi_{l+}^2 \\ \psi_{r-}^1 & \psi_{r-}^2  \end{array} \right|, \text{ etc.}$$
If we parametrize the unitary matrix $U \in U(2)$ as:
$$ U = e^{i\theta/2} \left[  \begin{array}{rc} \alpha & \beta \\ -\bar{\beta} & \bar{\alpha}  \end{array} \right], \quad | \alpha |^2 + | \beta |^2 = 1 ,$$
then the spectral function becomes:
\begin{eqnarray}\label{spectral_2}
\Lambda_U (\lambda) &=& W(l,r,-,-) +\alpha W(r,l,-,+)+\bar{\alpha}W(r,l,+,-)+\beta W(r,r,-,+)\\& &-\bar{\beta}W(l,l,+,-)+e^{i\theta} W(l,r,+,+)\notag
\end{eqnarray}
In particular if we consider a single interval $[0, 2\pi]$ with trivial riemannian metric,  the fundamental solutions to the equation Eq. (\ref{eigen_alpha}) have the form $\Psi^1 = e^{i\sqrt{2\lambda}x}$ and $\Psi^2 = e^{-i\sqrt{2\lambda}x}$.  Then we have:
\begin{eqnarray*} 
W(l,r,-,-) &=& -2i(1+2\lambda)\sin(2\pi\sqrt{2\lambda})-4\sqrt{2\lambda}\cos (2\pi\sqrt{2\lambda}), \\ 
W(l,l,+,-) &=& 4\sqrt{2\lambda}, \\
W(r,r,-,+) &=& 4\sqrt{2\lambda}, \\
W(r,l,-,+) &=& 2i(1-2\lambda)\sin(2\pi\sqrt{2\lambda}),\\
W(r,l,+,-) &=& 2i(1-2\lambda)\sin(2\pi\sqrt{2\lambda}),\\
W(l,r,+,+) &=& -2i(1+2\lambda)\sin(2\pi\sqrt{2\lambda})+4\sqrt{2\lambda}\cos (2\pi\sqrt{2\lambda}),
\end{eqnarray*}
and finally we obtain the spectral function $\Lambda_U(\lambda)$: 
\begin{eqnarray*} \Lambda_U( \lambda ) &=& -2i(1+2\lambda)\sin(2\pi\sqrt{2\lambda})-4\sqrt{2\lambda}\cos (2\pi\sqrt{2\lambda}) +4i\operatorname{Re}(\alpha) (1-2\lambda)\sin(2\pi\sqrt{2\lambda})\\
& &+  8\operatorname{Im}(\beta) \sqrt{2\lambda}+ e^{i\theta}[-2i(1+2\lambda)\sin(2\pi\sqrt{2\lambda})+4\sqrt{2\lambda}\cos (2\pi\sqrt{2\lambda})].
\end{eqnarray*}

\subsubsection{Quantum wires}

The discussion in the previous section allows to discuss a great variety of self--adjoint extensions
of 1D systems whose original configuration space  $\Omega = \sqcup_{\alpha=1}^n [a_\alpha, b_\alpha]$ consist of a disjoint union of closed intervals in $\mathbb{R}$.
It is clear that some boundary conditions $U\in U(2n)$ will lead to a quantum system with configuration space a 1D graph whose edges will be the boundary points $\{ a_1, b_1, \ldots, a_n,b_n\}$ of the original $\Omega$ identified among themselves according to $U$ and with links $[a_\alpha, b_\alpha]$. 

We will say that the self--adjoint extension determined by a unitary operator $U$ in $U(2n)$ defines a quantum wire made of the links $[a_\alpha, b_\alpha]$ if there exists a permutation $\sigma$ of $2n$ elements such that Asorey's condition for $U$ implies that $\psi (x_\alpha) = e^{i\beta_\alpha}\psi (x_{\sigma(\alpha)}$, and $x_\alpha$ such that $x_\alpha = a_\alpha$ if $\alpha = 1, \ldots, n$, or $x_\alpha = b_{\alpha-n}$ if $\alpha = n+1, \ldots, 2n$.

Notice that Asorey's condition:
$$\psi -i \dot{\psi} = U(\psi + i \dot{\psi}) $$
guarantees that the evolution of the quantum system is unitary, i.e., if we consider for instance a wave packet localized in some interval $[a_k,b_k]$ at a given time, after a while, the wave packet will have spread out accross the edges of the circuit, however the probability amplitudes will be preserved.  In this sense we may consider Asorey's equation above as the quantum analogue of Kirchhoff's circuit laws, or quantum Kirchhoff's laws for a quantum wire.



\section{Lecture 2:  Self--adjoint extensions of the covariant Laplace operator and the Hermitean Grassmannian}

\subsection{Von Neumann's theory of self--adjoint extensions and
boundary conditions}

In the previous lecture we have sketched a theory of
self--adjoint extensions of symmetric differential operators using as an example the Laplace--Beltrami operator, which is based on a geometrical structure, the Lagrange bilinear form.   
However a general (abstract) solution to this problem was given
by von Neumann \cite{Ne29}.    We will discuss now the exact nature
of the link between both approaches, the one based on geometrical boundary
data and von Neumann's theorem based on global information in the bulk.

It is interesting to point it out that there is a generalization of von Neumann's theory of
extensions of formally normal
operators with non-dense domains \cite{Co73}.  These results can be
discussed from the viewpoint of the geometry
of boundary conditions too.  We will not insist on this here and we
will restrict for clarity on the exposition to the simpler case
of self--adjoint extensions of symmetric operators with dense domains.

Suppose that $H$ denotes a symmetric operator on the Hilbert space $\mathcal{H}$ (for instance the Schr\"odinger operator Eq. \eqref{schrodinger}), then we may define the deficiency
spaces $\mathcal{N}_{\lambda}$, $\mathcal{N}_{\bar\lambda}$ for any $\lambda\in \C$,
$\mathrm{Im\,} \lambda\neq 0$, by setting,
\begin{equation}\label{deficiency} 
\mathcal{N}_{\lambda}  = \ran
( H + \lambda \mathbb{I})^\perp = \ker (H^\dagger + \bar\lambda \mathbb{I}),
~~~~ \mathcal{N}_{\bar\lambda}  = \ran ( H + \bar\lambda \mathbb{I})^\perp = \ker
(H^\dagger + \lambda \mathbb{I}) .
\end{equation}  
It is then true that for any complex $\lambda \notin \mathbb{R}$ the dimension of $\mathcal{N}_\lambda$ is constant on the upper (lower) half--plane and:
\begin{equation}\label{total_domain}
\mathcal{D} = \mathcal{D}_0 + \mathcal{N}_\lambda + \mathcal{N}_{\bar\lambda} ,
\end{equation}
and the sum is direct as vector spaces. 
Von Neumann's theorem \cite{Ne29} states that\footnote{Different
presentations of this theorem can be found in
\cite{Du63}, \cite{Na68} \cite{Yo65}, \cite{Re75} and \cite{We80}, existing
a vast literature on the subject.},

\begin{theorem}\label{von_Neumann} There exists a one-to-one correspondence between
self--adjoint (symmetric) extensions of $H$ and unitary operators (partial isometries) $K$ from
$\mathcal{N}_\lambda$ to $\mathcal{N}_{\bar\lambda}$, for any nonreal $\lambda\in \mathbb{C}$.
\end{theorem}

The domain $\mathcal{D}_K$ of the self--adjoint extension corresponding to the operator
$K$ is $\mathcal{D}_0 + \ran (\mathbb{I}+K)$.  The extension $H_K$ of the operator $H$ is defined for a function of the form
$\psi = \psi_0 + (\mathbb{I} + K) \xi_+$,
$\psi_0 \in \mathcal{D}_0$, $\xi_+\in \mathcal{N}_+$, by
$$ H_K \psi = H \psi_0 + \lambda \xi_+ + \bar{\lambda} K\xi_+
.$$ 

The main idea of the proof is to show that there is a one-to-one
correspondence between extensions of the symmetric operator $H$ and
extensions of its Cayley transform $U \colon \ran (H + \lambda \mathbb{I})
\to \ran (H + \bar\lambda \mathbb{I})$ defined by
$$ U = \frac{H + \bar\lambda \mathbb{I}}{H + \lambda \mathbb{I}} .$$ 

To compare with our previous results it will be convenient to 
describe von Neumann extension theorem in the
setting of skew-pseudo-hermitian spaces.

We define the total deficiency space $\mathcal{H}_{VN} =Ê\mathcal{N}_\lambda
\oplus \mathcal{ N}_{\bar\lambda}$.  As we have discovered in Thm. \ref{self2},
unitary operators from $\mathcal{ N}_\lambda$ to $\mathcal{ N}_{\bar\lambda}$ are in
one--to--one correspondence with maximal isotropic subspaces of $\mathcal{H}_{VN}$ with respect to the natural pseudohermitian structure
$\omega_{VN}$ defined on $\mathcal{ H}_{VN}$ by
\begin{equation}
\sigma_{VN} (\psi_1^+, \psi_1^-;\psi_2^+,\psi_2^-) = \langle \psi_1^+
,\psi_2^+ \rangle - \langle \psi_1^-,
\psi_2^- \rangle , ~~~~ \forall \psi_\alpha^+ \in \mathcal{N}_\lambda, \psi_\alpha^-
\in \mathcal{N}_{\bar\lambda} .
\end{equation}
Now we can try to identify the total deficiency space
$\mathcal{H}_{VN}$ with the space of boundary data defined by the Laplace--Beltrami operator.

Before doing this it will be convenient to enlarge slightly our setting by allowing differential operators action on spaces of vector--valued functions, i.e., sections of a vector bundle, instead of scalar functions.   Thus the Laplace--Beltrami operator will be replaced by the covariant Laplacian $\Delta_A$.     Following closely the notations and conventions in Section \ref{Laplace_Beltrami} we may consider $\pi \colon E \to \Omega$ be an Hermitean bundle over $\Omega$
of rank $r$, whose Hermitean structure will be denoted by $(\cdot,\cdot )$.   
We will denote as well by $\Gamma^\infty (E)$ the space of smooth sections of the bundle $E$ and by $\Gamma_c^\infty (E)$ the space of smooth sections with support on the interior of $\Omega$.  

A Hermitean connection $\nabla$ on the bundle $E$ is by definition a linear differential operator $\nabla \colon \mathfrak{X}(\Omega) \times\Gamma^\infty (E) \to \Gamma^\infty (E)$
such that:  i) $\nabla_X(f \Phi) = f \nabla_X\Phi + X(f) \Phi$, ii)  $(\nabla_X \Phi,\Psi) +  (\Phi,\nabla_X \Psi) = X(\Psi, \Psi)$ for all $X\in \mathfrak{X}(\Omega)$, $\Phi, \Psi \in \Gamma^\infty (E)$, $f \in C^\infty(\Omega)$.

We will denote by  $\mathcal{H}^k(E)$ the Hilbert space of equivalence classes of sections of the bundle $E$ of Sobolev class $k$, i.e.,  a section $\Phi \in \Gamma^\infty(E)$ is of Sobolev class $k$ if $|| \Phi ||_k^2 =  \int_\Omega (\Phi (x), (I - \nabla_0^\dagger \nabla_0 )^{k/2}\Phi(x))_x \vol_\eta (x) < \infty$ where $\nabla_0$ is a fixed reference Hermitean connection on $E$ and $\nabla_0^\dagger$ is the formal adjoint differential operator of $\nabla_0$.  Then $\mathcal{H}^k(E) = \overline{\Gamma^\infty(E)}^{||\cdot ||_k}$. 

The restriction of the bundle $E$ to the boundary $\partial \Omega$, again denoted by $\Gamma$ in what follows,  will be denoted by $\partial E$, i.e., $\partial E = E\mid_{\Gamma}$, and the restriction of the projection $\pi$ to $\partial E$, by $\partial \pi$, thus $\partial \pi \colon \partial E \to \Gamma$ is again an Hermitean bundle over $\Gamma$ of rank $r$.   Any Hermitean connection $\nabla$ of $E$ restricts to an Hermitean connection of $\partial E$ that will be denoted with the same symbol.  Thus the space of smooth sections of $E$ restricted to $\Gamma = \partial \Omega$ is $\Gamma^\infty(\partial E)$.

We will consider the Bochner Laplacian (that will be also called the covariant Laplacian) associated to the Hermitean connection $\nabla$ as the formally self--adjoint elliptic differential operator of order 2 acting on sections of the Hermitean bundle $E$ with compact support on $\Omega \backslash \partial\Omega$ by  $\Delta_A = -\nabla^\dagger \nabla$.    We will denote again by $\Delta_0$ the minimal closed extension of the operator $\Delta_A$ with respect to the graph--operator norm as in Section \ref{Laplace_Beltrami}.  It is well--known that the domain of $\Delta_0$ is given again by $\mathcal{D}( \Delta_0)= \mathcal{H}_0^2(E)$.   We will denote by $\Delta_0^\dagger$ the adjoint operator of $\Delta_0$ in $L^2(E)$ whose domain contains $\mathcal{H}^2(E)$.    

By $\varphi:=\Phi|_{\partial\Omega}$ and $\dot{\varphi}:=\nabla_\nu\Phi|_{\partial\Omega}$ we denote
again the restriction of $\Phi \in \mathcal{H}^2(E)$ to the boundary 
and the covariant normal derivative with respect to the outward normal respectively.
We will call the pair $(\varphi, \dot{\varphi})$ the boundary data of $\Phi$ and we will denote it by $b (\Phi)$.

The induced scalar product
on the boundary is denoted again by 
$$
\langle \varphi ,\psi \rangle = \int_{\Gamma }(\varphi (x), \psi(x))_x \,\vol_{\partial \eta}(x)  .
$$ 

The boundary map $b \colon \Gamma^\infty (E) \to \Gamma^\infty(\partial E) \times \Gamma^\infty(\partial E)$ can be extended continuously to $\mathcal{H}^2(E)$, which constitutes another statement of the well-know Lions-Mag\'enes trace theorem, Thm. \ref{trace}.   In this context the weak trace theorem for the Bochner Laplacian states that  there is a unique continuous extension of the boundary map $b$ such that $b \colon \mathcal{H}^2(E) \to \mathcal{H}^{3/2}(\partial E) \oplus \mathcal{H}^{1/2}(\partial E)$.  Moreover the map is surjective and $\ker b = \mathcal{H}_0^2(E)$.  We will denote by $\mathcal{H}_L = \mathcal{H}^{3/2}(\partial E) \oplus \mathcal{H}^{1/2}(\partial E)$ the Hilbert space of boundary data $(\varphi, \dot{\varphi})$.

We will assume in what follows that the self--adjoint extensions of the Bochner Laplacian we are interested in are such that the graph of the unitary operator $K\colon \CN_+ \to \CN_-$ is contained in $\mathcal{H}^2(E)$. Then the boundary map $b$ restricts to the graph of $K$ which is contained in $\mathcal{H}_{VN}\cap \mathcal{H}^2 (E)$.  
We compose $b$ with the Cayley transform on the
boundary $C$ to obtain a continuous linear map $j\colon \mathcal{H}_{VN}\cap \mathcal{H}^2 (E)
\to \mathcal{H}_L$ defined as follows.  Let $j_\pm (\psi^\pm) = \varphi \pm i
\dot{\varphi}$, where $(\varphi, \dot\varphi) = b (\psi )$, and
we will denote $j_\pm (\psi^\pm )$ as usual by $\varphi^\pm$.  Then, $j =
j_+ \oplus j_-$ or explicitly,
\begin{equation}\label{j}  
j(\psi^+, \psi^-) = ( \varphi^+, \varphi^- ) .
\end{equation}

The following lemmas will
show that $\, j$ is an isometry of skew-pseudo-hermitian structures.

\begin{lemma}  With the notation above the map $j\colon \mathcal{H}_{VN}
\to \mathcal{H}_L$ verifies
$$ \sigma_{VN} (\psi_1^+, \psi_1^-;\psi_2^+,\psi_2^-) = \Sigma
(\varphi_1^+, \varphi_1^-;\varphi_2^+,\varphi_2^-) .$$ 
\end{lemma}

{\it Proof:}  We consider $\lambda = i$, the proof for general
$\lambda$ proceeds equally. 
We shall consider first $\psi_1^+, \psi_2^+ \in \mathcal{N}_i$, then
$-\Delta_A^\dagger \psi_a^+ = i\psi_a^+$, $a=1,2$.
Then it is clear that,
\begin{eqnarray*}
0 &=& \langle \psi_1^+, (-\Delta_A^\dagger - i)\psi_2^+ \rangle =
\langle \psi_1^+, -\Delta_A \psi_2^+ \rangle -i \langle \psi_1^+,
\psi_2^+ \rangle \\
&=& \langle -\Delta_A\psi_1^+, \psi_2^+ \rangle -i \Sigma_B
(b(\psi_1^+),b(\psi_2^+)) - i \langle \psi_1^+, \psi_2^+ \rangle \\
&=& \langle (-\Delta_A -i)\psi_1^+, \psi_2^+ \rangle - 2i \langle
\psi_1^+, \psi_2^+ \rangle - i\Sigma
(b(\psi_1^+),b(\psi_2^+)) \\  &=& - 2i \langle
\psi_1^+, \psi_2^+ \rangle - i\Sigma (b(\psi_1^+),b(\psi_2^+)).
\end{eqnarray*}
Hence,
$$\sigma_{VN} (\psi_1^+,0;\psi_2^+,0) = \langle \psi_1^+, \psi_2^+ \rangle
= -\frac{1}{2} \Sigma (b(\psi_1^+),b(\psi_2^+)) = -\frac{1}{2}
\Sigma (\varphi_1^+,0; \varphi_2^+,0) .$$
Similarly, it is shown that  $\sigma_{VN} (0,\psi_1^-;0,\psi_2^-) = \Sigma
(0,\varphi_1-;0,\varphi_2^-)$. \hfill $\Box$

\bigskip

To show that $j$ is onto we will
need the following result about from the existence and uniqueness of solutions of the Dirichlet
problem.

\begin{proposition}\label{existence} For every $\phi \in \Gamma (\partial
E)$, and for every non real $\lambda$ there is a unique solution of the
equations
\begin{equation}\label{delta_i}  
- \Delta_A \Psi + \lambda \Psi = 0 , 
~~~~ - \Delta_A \Psi + \bar\lambda \Psi = 0 
\end{equation}  
with boundary condition
$$\Psi \mid_{\partial \Omega} = \phi .$$
\end{proposition}

{\it Proof:}   We prove first uniqueness.  If there were two different
solutions
$\Psi_1$, $\Psi_2$, then because the operator $-\Delta_A + \lambda$ is
elliptic, by elliptic regularity they will be both smooth.   Then, $\Psi =
\Psi_1 - \Psi_2$ also satisfies Eq. (\ref{delta_i}) with boundary
condition
$$\Psi \mid_{\partial \Omega} = 0 ,$$  
which is impossible by the uniqueness of the solution of the Dirichlet problem.  In fact, if we look
for solutions $\Psi$ of the equation (\ref{delta_i}) such that
$\Psi \mid_{\partial \Omega} =$ constant, then, we can remove the boundary
identifying all their points and looking for the solutions of eq.
(\ref{delta_i}) on the closed manifold $\Omega^\prime$ obtained in this
way.  But now, $-\Delta_A$ is essentially self--adjoint on
$\Gamma (E^\prime)$ where $E^\prime$ is the fibre bundle obtained from $E$
identifying all the fibres over $\partial \Omega$\footnote{Notice that
the compactness of $\Omega$ is crucial in this statement.}, and then it has
not imaginary eigenvalues.   

\medskip

Let us now prove the existence of solutions.  Let $\tilde{\Psi}$ be any
section in $\Gamma (E )$ such that $\tilde{\Psi}
\mid_{\partial \Omega} = \phi$.  Then, there exists a unique section
$\zeta \in \Gamma (\Omega )$ such that
$$ - \Delta_A \zeta + \lambda \zeta = \Delta_A \tilde{\Psi} - \lambda
\tilde{\Psi} ,$$  
with Dirichlet boundary conditions. This is a consequence of the
solution of the Dirichlet boundary value problem for elliptic
operators.  Then, the section $\Psi = \zeta + \tilde{\Psi}$ verifies Eq.
(\ref{delta_i}) and the boundary condition $\Psi \mid_{\partial
\Omega} = \phi$. 
\hfill $\Box$

\bigskip

\begin{theorem}\label{ident} The deficiency space on the bulk $\mathcal{H}_{VN}\cap \mathcal{H}^2 (E)$ with its natural skew--Hermitean structure $\sigma_{VN}$ is
isometrically isomorphic to the boundary data space $\mathcal{H}_L$ with its
natural skew--Hermitean structure $\Sigma$.
\end{theorem}

{\parindent 0cm{\it Proof:}}  We will have to show that the map $j$ is
onto. We can solve the boundary problems
\begin{eqnarray}   -\Delta_A \Psi^+ + \lambda \Psi^+ = 0 ,&& \quad 
\Psi^+\mid_{\partial
\Omega} = \varphi^+ \\ -\Delta_A \Psi^- + \bar\lambda \Psi^- = 0 ,&& \quad
\Psi^-\mid_{\partial \Omega} = \varphi^- , 
\end{eqnarray} 
for given $\varphi^\pm \in \Gamma^\infty (\partial E)$. 
Proposition
\ref{existence} shows that such solutions $\Psi^\pm$ exist and
they are unique.  They define the inverse of the map $j$ on the dense
subspace $\Gamma^\infty (\partial E)\oplus \Gamma^\infty (\partial E)$, thus $j$ is an isometry onto. 
\hfill $\Box$

\bigskip

Notice that the previous theorem can also be seen as offering an
alternative proof of von Neumann's theorem for the symmetric operator $\Delta_A$.  
Similar arguments can be reproduced in the much broader context of symmetric pseudodifferential operators of
any order in compact manifolds with boundary.   For instance the results obtained so far can be used to obtain a similar theory for Dirac operators.  We will come back to this in Lecture 3.


\subsection{Self--adjoint extensions, boundary data and
Cayley submanifolds}

The characterization of self--adjoint extensions of
$H = -\Delta_0$ in terms of a class of unitary operators in $\mathcal{U}(L^2(\Gamma, \mathbb{C}^r))$\footnote{Notice that as Hilbert spaces $L^2(\Gamma, \C^r)$ is the same as the Hilbert space $L^2(\partial E)$ of square integrable sections of the restriction of the bundle $E$ to the boundary.}, although
similar to von Neumann characterization,
is more useful for applications because it is formulated in terms of boundary data.    The constraints
involved in the definition of the domain determined by the unitary operator $U$ imply that the
boundary values $\varphi$, $\dot\varphi$ of the functions of
such a domain satisfy Asorey's condition Eq. \eqref{asorey}.
Generically,  Eq. \eqref{asorey} can be solved to express
$\dot\varphi$ as a function of $\varphi$, i.e.,

\begin{equation}  \label{phidotphi}
\dot\varphi =- i \frac{\mathbb{I}-U}{\mathbb{I}+U} \varphi
\end{equation}
or, alternatively,
$\varphi$ as a functions of $\dot\varphi$
\begin{equation}\label{phiphidot}
\varphi = i \frac{\mathbb{I} + U}{\mathbb{I} - U}  \dot\varphi .
\end{equation}
Notice that a necessary and sufficient condition for the existene of $(\mathbb{I}\pm U)^{-1}$ is that $\mp 1$ is not in the spectrum of $U$ respectively.
This explicit resolution of the constraint on the boundary data
means that unitarity requires that only half of the dynamical data
are independent at the boundary.

Equations \eqref{phidotphi} and   \eqref{phiphidot}  are in fact two different expressions of the
Cayley transform relating self--adjoint  and unitary operators:
\begin{equation}\label{cayley}
A=-i \frac{\mathbb{I}-U}{ \mathbb{I}+U}; \qquad A^{-1} = i \frac{\mathbb{I} + U}{ \mathbb{I} - U}.
\end{equation}
The inverse transformation being also a Cayley transform
\begin{equation} \label{invcayley}
U= \frac{\mathbb{I} - iA}{\mathbb{I} + i A} \, .
\end{equation}

Notice that contrary to what happens with the definition of $A$ in terms of $U$, given a self--adjoint operator $A$, the unitary operator $U$ given by Eq. \eqref{invcayley} is always well--defined.

These considerations show that there is a distinguished set of
self--adjoint extensions of $H$ for which the expression of the
boundary conditions defining their domain cannot be reduced to the
simple form given by Eqs. \eqref{phidotphi} or  \eqref{phiphidot}.  These
self--adjoint extensions correspond to the cases where $\pm 1$ are in the
spectrum of the corresponding unitary operator $U$.

The Cayley subspaces $\mathcal{C}_\pm $
are thus defined as the subspaces of self--adjoint extensions
which cannot be defined in the form \eqref{phidotphi} or \eqref{phiphidot}, i.e.:
\begin{equation}\label{cayleysur} 
\mathcal{C}_\pm = \left\{U\in \mathcal{U}\left( L^2 (\Gamma,\C^r)\right)\Big| \pm 1 \in \sigma (U)\right\} \, .
\end{equation}
Notice that the unitary operators $U=\pm \mathbb{I}$  are in the Cayley subspaces
$\mathcal{C}_\pm$, respectively.  $U=-\mathbb{I}$ belongs to the Cayley subspace
$\mathcal{C}_-$ and corresponds to
Dirichlet boundary conditions:
\begin{equation}\label{dirich}  \varphi = 0 ,
\end{equation}
whereas  $U=\mathbb{I}$ is in the Cayley subspace
$\mathcal{C}_+$ and corresponds to the self--adjoint
operator $A = 0$ which defines Neumann boundary conditions
\begin{equation}\label{neuma}  \dot\varphi = 0 .
\end{equation}

There is a formal property which distinguishes the two Cayley subspaces. The
subspace $\CC_+$ has a group structure whereas $\CC_-$ does not because the composition is not a inner operation.  Notice that neither
$\CC_-\cap\,\CC_+$ has a group structure.

We will denote by $\CM$ the space of self--adjoint extensions of the Bochner Laplacian 
$\Delta_A$.  Notice that so far we have described a family of self--adjoint extensions characterized by the property that their domains are contained in $\mathcal{H}^2(E)$.  In fact because of Thm. \ref{self1} such extensions are in one--to--one correspondence with the subgroup of the group of unitary operators $U \in \CU(L^2(\partial E))$ preserving the subspace $\mathcal{H}^{3/2}(\partial E) \oplus \mathcal{H}^{1/2}(\partial E) \subset L^2(\partial E) \oplus L^2(\partial E)$, i.e., the unitary operators $U$ such that $C^{-1}(\mathrm{graph}(U) \cap \mathcal{H}^{3/2}(\partial E) \oplus \mathcal{H}^{1/2}(\partial E)$ is a closed maximally isotropic subspace.
Thus the identification of the space $\CM$
with a subgroup of the unitary group $\CU(L^2(\Gamma,\C^r))$
provides an explicit group structure to this space of
self--adjoint realizations of $\Delta_A$.

In what follows we will identify this space of self--adjoint extensions of the Bochner Laplacian 
$\Delta_A$ with the unitary group $\CU(L^2(\Gamma, \mathbb{C}^r))$ itself because it can be proved that it provides a parametrization of all self--adjoint extensions of $\Delta_A$ (see \cite{Ib12} and references therein) and we will denote it by $\CM$ again.

\subsection{The self--adjoint Grassmannian}\label{selfadjoint_grassmannian}

The space $\CM$ of self--adjoint extensions of the Bochner Laplacian
has a non-trivial topological structure.
All even homotopy groups vanish
$\pi_{2n}(\CM)=0$ but all odd homotopy
groups are non-trivial $\pi_{2n+1}(\CM)=\Z$ because of Bott's periodicity theorem.
The fact that the first homotopy group
$\pi_1(\CM)=\Z$ is non--trivial means that the space of boundary conditions is
non-simply connected.  However the set of self--adjoint operators
in  $L^2(\Gamma, \C^r)$ is a topologically trivial manifold (notice that any self--adjoint operator $A$ can be deformed homotopically to 0 by $(1-t)A$, $t \in [0,1]$).  This means that
the characterization of self--adjoint extensions of $\Delta_A$
by means of the Cayley transform \eqref{phidotphi} and \eqref{phiphidot}  cannot provide
a global description of $\CM$. In fact, the parametrization
\eqref{invcayley} and its inverse
 \begin{equation}\label{invcayleyy} U^{-1}= \frac{\mathbb{ I} + iA}{ \mathbb{I} - iA}.
 \end{equation}
can be considered as local coordinates in the  charts $\CM \setminus  \CC_\pm$
of the space $\CM$  of self--adjoint  extensions $\Delta_A$.
The topology of each chart is trivial but that of $\CM$ is
not. In this sense, the Cayley submanifold $\CC_\pm$
intersects all non-contractible cycles of $\CM$. 

Since $\pi_0 (\CM ) = 0$ and $\pi_1(\CM) = \Z$ the first cohomology group
of $\CM$ is $H^1(\CM)=\Z$. The generator of this cohomology group
is given by the first Chern class of the determinant bundle defined over 
$\CM$. The determinant of infinite dimensional unitary operators $U$ is ill
defined and its proper definition requires the introduction of a regularization.
In particular, it is  necessary to restict the boundary conditions
to the subspace $\mathcal{M}'$ defined by the  unitary $U$ operators of
$\mathcal{M}$ which are of the form $U=\mathbb{I} + K$ with $K$ a  Hilbert--Schmidt
operator (i.e. $ \Tr (K^\dagger K) <  \infty$).
 If $-1\notin\sigma (U)$ this property
 is equivalent to the requirement that the Cayley transform of the
operator $A$ is also Hilbert--Schmidt.  Indeed, 
$$
A= \frac{i K} {2 \mathbb{I} + K}, \qquad K = \frac{2A}{i\mathbb{I} - A}; 
$$
hence,
$$
K^\dagger K = \frac{4A^2}{\mathbb{I} + A^2};  \qquad A^\dagger A= \frac{K^\dagger K}{ (2 \mathbb{I} + K^\dagger) (2 \mathbb{I} + K)}
$$
and we get the bounds:
$$ 
\Tr ( K^\dagger K) = 4\, \Tr  \left(\frac{A^2}
{\mathbb{I} + A^2} \right) \leq 4\, \Tr  ( A^2) , \quad \Tr ( A^\dagger A ) \leq  
\Tr  ( K^\dagger K) .
$$

With this restriction the determinant of $U \in \CM'$ can be defined 
by using the standard renormalization
prescription for  determinants
$$
\log \left( {\det}' \, U \right)=  \sum_{k=1}^\infty d_k e^{-\lambda_k}\,\log\  (1+\lambda_k) ,
$$
in terms of the eigenvalues of $K$, $\lambda_k$, $k=1,2,\cdots $,
and their degeneracies $d_k$, $k=1,2,\cdots$.
Finiteness of this prescription for the
 regularized determinant  ${\det} ' \, U$ follows from the Hilbert-Schmidt
character of $K$ which in particular implies a
discrete spectrum with finite degeneracies 
satisfying the Hilbert-Schmidt condition 
$K^\dagger K = \sum_{k=1}^\infty d_k |\lambda_k|^2\leq \infty$.

The first Chern class of the regularized determinant bundle
is given by the one--form:
\begin{equation}\label{chc}
\alpha= \frac{1}{2\pi } \, \mathrm{d} \left(\log \left({\det} ' \, \gamma(\theta) \right)\right)\, .
\end{equation}
For any closed curve $\gamma \colon S^1 \to \mathcal{M}'$
in the self--adjoint grassmannian,
we  define its  Maslov index $\nu_M(\gamma )$ as the winding
number of the curve ${\det}' \circ \gamma \colon S^1 \to U(1)$  (see for instance \cite{Ar67}).
In other words,
\begin{equation}\label{maslov}
\nu_M (\gamma) = \frac{1}{2\pi} \int_0^{2\pi}\partial_\theta
\,\log\,\left( {\det}' \, \gamma
(\theta) \right) \, d\theta.
\end{equation}
Thus the Maslow index $\nu_M(\gamma )$ is the sum of the winding numbers
of the maps 
$\lambda_i(\theta):S^1\to U(1)$ 
described by the flow of eigenvalues of $\gamma$
around $U(1)$. By continuity of $\gamma$ and compactness of $S^1$ it 
follows that only a finite number of eigenvalues reach the value 
$\lambda_i=-1$ for any value of $\theta\in [0, 2\pi)$. It is clear 
that the winding number of the  map $\lambda_i(\theta)$ is
measured by  $\frac{1}{2\pi} \int_0^{2\pi}\partial_\theta
\log\,  (\lambda_i
(\theta)) d\theta$ and also by
the number of indexed crossings of the point $\lambda_i=-1$.
By construction $\nu_M (\gamma)$ is  the finite sum of the 
non-trivial  winding numbers and is always an integer.
This fact and the  existence of  curves with only one crossing
through $-1$ implies that $\alpha$ is in  the generating class of 
the cohomology group
$H^1(\CM',\IZ)$.

The subspace $\CM'$ of unitary operators of the form $U=\mathbb{I} + K$
has richer topological and geometrical structures.  In particular
we will see that it is a Grassmaniann, the self--adjoint Grassmannian as it will be called in
what follows.

It is obvious that the subspaces
$\CH_+ = L^2 (\Gamma,\C^r) \times \{ {\bf 0} \}$ $= \set{(\varphi, {\bf
0}) \mid \varphi \in L^2 (\Gamma,\C^r)}$ and $\CH_- = \{ {\bf 0}
\} \times L^2 (\Gamma,\C^r)$ $= \set{({\bf 0}, \dot\varphi) \mid
\dot\varphi \in L^2 (\Gamma,\C^r)}$, which correspond to
Dirichlet and Neumann boundary conditions respectively,
are isotropic in $\mathcal{H} = L^2(\Gamma, \mathbb{C}^r) \oplus L^2(\Gamma, \mathbb{C}^r)$ and they are paired by
$\Sigma$.  In fact,
$$\Sigma (\varphi_1,{\bf 0}; {\bf 0},\dot{\varphi}_2) = 
\langle \varphi_1, \dot{\varphi}_2 \rangle_{ \Gamma}  $$
The block structure of $\Sigma$ with respect to the isotropic
polarization $ \CH_+ \oplus \CH_-$ of $\CH$ reads
$$\Sigma = \left[ \begin{array}{c|c} 0 & \langle
\cdot \,, \cdot \,\rangle_{\Gamma} \\ \hline
 - \langle \cdot \,, \cdot \,\rangle_{\Gamma} & 0
\end{array} \right] .$$
 
The pseudo--Hemitean   structure $\Sigma$ can be diagonized by means of the Cayley transform
\begin{equation}\label{cayley_transform} 
C (\varphi, \dot\varphi ) =(\phi^+, \phi^- )=
\left( \frac{1}{\sqrt{2}}(\varphi + i \dot\varphi ), \frac{1}{\sqrt{2}}(\varphi - i \dot\varphi) \right).
\end{equation}
which transforms $\Sigma$ into 
$\widetilde\Sigma$.

There is  another canonical hermitian 
product on $ \CH_+ \oplus \CH_-$
given by the matrix operator:
$$ \left[\begin{array}{c|c} \mathbb{I} & 0  \\ \hline
0 & \mathbb{I} \end{array}
\right] $$
which defines a Hilbert structure $\langle \cdot \, , \cdot \,\rangle$ 
on $ \CH_+ \oplus \CH_-$.

The  Grassmannian  $\mathrm{Gr} (\CH_+,\CH_-)$ of $ L^2 (\Gamma, \C^r)
\times L^2 (\Gamma,\C^r)$  is the infinite--dimensional Hilbert
manifold of closed subspaces $W$ in $ \CH_+ \oplus \CH_-$ such that
the projection on the first factor $\pi_+ \colon W \to \CH_+$ is
a Fredholm operator and the projection on the second factor $\pi_- \colon  W
\to \CH_-$ is Hilbert--Schmidt, that is, $\Tr  (\pi_-^\dagger\pi_- )< \infty$.

The  self--adjoint Grassmaniann  $\mathrm{Gr} (\CH_+,\CH_-)\cap\,\CM$
is defined by the self--adjoint extensions of $\Delta_A$ which belong
to the Grassmaniann $\mathrm{Gr} (\CH_+,\CH_-)$.  This subspace might be
considered as the space of ``mild'' self--adjoint extensions of $\Delta_A$.
It is possible to see that the self--adjoint Grassmannian is an open
submanifold of the Grassmannian itself and can be identified with $\CM'$,
the space of unitary operators of $\CM$ which are of the 
form $U=\mathbb{I} + K$. This follows from the fact that in the previous parametrization
of $\CM'$ we have:
\begin{equation}\label{marv} \pi_-=\frac{iK}{2\sqrt{U}},
\end{equation}
which implies that 
$\Tr  (\pi_-^\dagger \pi_- ) = \frac{1}{4}\Tr  (K^\dagger K)$, i.e.
 $\pi_-$ is Hilbert--Schmidt if and only if $  K$ is Hilbert--Schmidt.

The intersection of the Cayley submanifold $\CC_\pm$
with $\CM'$ defines a subspace of the self--adjoint
Grassmannian $\CC'_\pm \subset \CM'$  which has a stratified structure
according to the number of eingenvalues $\pm 1$ of the corresponding
unitary operator, i.e.
$$\CC'_\pm=\bigcup_{n=1}^{\infty} \CC^{'n}_\pm,$$
 where
$\CC^{'n}_\pm=\{U\in \CU(L^2(\Gamma,\C^r)\mid \pm 1 \in  \sigma (U)\
{\rm with\  multiplicity}\  n\}$.
Notice that the  spectrum of unitary operators in the self--adjoint
Grassmannian is discrete.

Given a continuous curve $\gamma \colon [0,1] \to \CM'$ we 
define its Cayley index $\nu_c(\gamma)$ as the indexed sum of 
crossings of $\gamma\,$ through the Cayley submanifold $\CC'_-$ 
(notice that the Cayley
submanifolds $\CC'_\pm$ are
oriented).
This is equivalent to the
sum of anti-clockwise  crossing of eigenvalues of
$\gamma\,$ through the point $-1$ on the unit circle
$U(1)$ minus the sum of clockwise crossings weighted with
the respective degeneracies. Therefore, the Cayley index
$\nu_c(\gamma)$  of $\gamma$  is 
equivalent to its  Maslow index $\nu_M(\gamma)$ and we have the
following theorem.

\begin{theorem}\label{maslov_index} The Maslov and  Cayley indices of a closed curve $\gamma$ in
the self--adjoint Grassmannian coincide, $\nu_M(\gamma) = \nu_c(\gamma)$.
Thus the Cayley manifold $\CC'_-$ is dual of the Maslov class $\alpha$.
\end{theorem}

For any  unitary operator $U\in \CM$
we will define its degenerate dimension as
the dimension of the eigenspace with eigenvalue $-1$.   If $U$ is in
the self--adjoint Grassmaniann $\CM'$ the dimension of the eigenspace  
with  eigenvalue $-1$ is finite and the
degenerate dimension of the operator is finite.  We shall denote such
number by $n(U)$ and it is an indicator of the level of $\gamma(\theta)$
in the stratified structure 
of $\CC'$: $U=\gamma(\theta)\in\CC^{'n}$ if and only if $n (U)=n$.
 The Cayley index of  any curve $\gamma\in \CM'$ 
can be given in terms of this number by the expression
\begin{equation}\label{rnu}
\nu_c(\gamma ) =\frac{1}{2\pi} \int_0^{2\pi}\partial_\theta
n (\gamma(\theta) ) d\theta .
\end{equation}
Since the r.h.s. of Eq. \eqref{rnu} is the integral of a pure derivative it vanishes unless there
is a singularity in the integrand. This only occurs at the jumps of
$n(\gamma(\theta))$ i.e.  when one more eigenvalue of 
$U=\gamma(\theta)$ becomes equal to $-1$. $\nu_M(\gamma)$ is in fact
a bookkeeping of the number of eigenvalues of $\gamma(\theta)$ that cross 
through $-1$ and since it is of bounded variation on  
$\CM'$  the integral in Eq. \eqref{rnu} is always finite and gives the Cayley
index.  This construction provides an alternative 
(singular) characterization of the first Chern
class of the determinant bundle $\det_{\CM'}\left(\CM',U(1)\right)$
and the generating class of the first homology group $H^1 (\CM', \Z)$ of 
$\CM'$. 

\subsection{Topology change and edge states.}

Although the operator $\Delta_A+\mathbb{I}$ is positive in $\Gamma^\infty_0(E)$,
its self--adjoint extensions might not be.
In fact, if the self--adjoint extension does not belong to any of the Cayley
submanifolds $\CC_\pm$ it is easy to show by integration by parts that
\begin{eqnarray} \langle\nabla \Psi_1,\nabla  \Psi_2\rangle &=& \langle\Psi_1,\Delta_0
\Psi_2\rangle + \langle\varphi_1, \dot\varphi_2\rangle \\ &=&
\langle\Psi_1, \Delta_0\Psi_2\rangle +  \langle\varphi_1, A
\varphi_2\rangle =  \langle\Psi_1,\Delta_0 \Psi_2\rangle +
\langle A^{-1} \dot\varphi_1, \dot\varphi_2\rangle ,
\end{eqnarray}
where $A =C(U)$ is the Cayley transform of the unitary operator defining the self--adjoint extension,
$\langle \nabla \Psi, \nabla \Phi \rangle = \int_\Omega \eta^{ij}(x) h_{ab}(x) \nabla_i \bar{\Psi}^a(x) \nabla_j \Phi^b (x)\vol_\eta(x)$, with $\Phi = \Phi^a \sigma_a$, with $\sigma_a$ and local reference frame, $h_{ab}(x) = (\sigma_a, \sigma_b)_x$, and $\nabla_i = \nabla_{\partial/\partial x^i}$.

Thus, only if  $\langle \nabla \Psi,\nabla \Psi\rangle-\langle \varphi, A
\varphi\rangle$ is positive for every $\Psi$, the operator
$\Delta_U$ will be positive.
In particular if the boundary operator $A$ is positive it might occur
that the whole operator $\Delta_U$ might loose positivity.
The existence of negative energy levels is thus possible for some
boundary conditions.  It can be seen that the states which have
negative energy are related to edge states as it is illustrated by the following result.

\begin{theorem}\label{edge_states}  For any self--adjoint extension  $\Delta_U$  of 
$\Delta_0$ whose unitary operator $U$ has one eigenvalue $-1$ with smooth eigenfunction,
the family of self--adjoint extensions of the form $U_t = U {\rm e}^{i t}$ 
with $t\in (0,\pi/2)$, has for small values of $t$, one negative energy level which corresponds to an 
edge state. The energy of this edge state becomes infinite when $t\to 0$.
\end{theorem}

{\parindent 0cm \emph{Proof:}}
Let  $\xi \in L^2(\Gamma,\C^r)$ be a smooth eigenstate
of $U$ with eigenvalue $-1$. Then, $U_t\xi={\rm e }^{i t} \xi$.
Let us consider Gaussian coordinates in a collar $\CC_\Gamma\subset \Omega$
around the boundary $\Gamma$ of $\Omega$.   One of those coordinates is the
``radius'' $r$ and the others can identified with boundary
coordinates sifted inside the collar; i.e. $\CC_\Gamma \approx
[1-\epsilon,1]\times \Gamma$.   In these coordinates the metric matrix
looks like
\begin{equation}\label{gmet} 
\eta = \left[\begin{array}{cc}1 & 0 \\  0 & \Lambda(r,\Gamma) \end{array} \right].
\end{equation}
We consider now the following change of coordinates
$r\mapsto s$ with $s= \frac{\pi}{2 \epsilon}(1-r)$.

If we extend the function $\xi$ from the boundary $\Gamma$ to an edge
state $\Psi$ in the bulk $\Omega$ by
\begin{equation}\label{ext}
\Psi(x)= \left\{ \begin{array}{l} \xi(\Gamma) e^{-k\tan s},
\quad x=(s,\Gamma)\in \CC_\Gamma  \\
0, \quad  x\notin \CC_\Gamma \end{array}
\right. ,
\end{equation}
it is easy to check that the extended  function $\Psi$ is smooth in $\Omega$
and for
$$k= \frac{2\epsilon}{\pi} \cot \frac{t}{2}$$
belongs to the domain of the self--adjoint extension of
$\Delta_{U_t}$ associated to the unitary matrix $U_t=
{\rm e}^{i t} U$.
Thus, we have
\begin{equation}\label{bcc}
\langle\Psi,\Delta_{U_t} \Psi\rangle = \langle \nabla \Psi, \nabla \Psi\rangle -  \cot\frac{t}{2}
\langle \xi, \xi \rangle
\end{equation}
where
\begin{eqnarray}\label{dd}
\langle \nabla \Psi, \nabla \Psi\rangle &=& \int_{0}^{\pi/2}
 ds\ \int_\Gamma\
d\mu_\Gamma(s) \  |\xi |^2  \left(\frac{k^2\pi}{2\epsilon}\right)
\left(1+(\tan s)^2\right)^2{\rm e}^{-2k\tan s} \\ \nonumber
&+& \frac{2 \epsilon}{\pi}\int_{0}^{\pi/2}  ds \ \int_\Gamma\
d\mu_\Gamma(s)\ \langle \xi^\ast ,\Delta_\Gamma \xi\rangle\  {\rm e}^{-2k\tan s}.
\end{eqnarray}
For small enough $\epsilon \ll1$ we have that the dependence
on $s$ of $\Lambda(s,\Gamma )$ might be negligible $|\Lambda(s,\Gamma)| <
|\Lambda(0,\Gamma)|(1 + \delta) $. Thus,
\begin{equation}\label{dzd}
\langle \nabla \Psi, \nabla \Psi\rangle \leq \left( \frac{k}{2} + 
\frac{1}{4 k} \right) \, \frac{\pi(1 + \delta)}{ 2\epsilon} \, \|\xi\|^2 +
\frac{2 \epsilon (1+ \delta)}{\pi } \langle \xi^\ast , \Delta_\Gamma
\xi \rangle ,
\end{equation}
and
\begin{equation}\label{bbc}
\langle \Psi,\Delta_{U_t}
\Psi\rangle \leq \frac{\pi}{2\epsilon}\left(  \frac{1}{4 k} (1+ \delta)-
\frac{k}{2}\, (1- \delta)\right)\,
\|\xi\|^2 + \frac{2 \epsilon  (1+ \delta)}{\pi} \langle \xi^\ast,
\Delta_\Gamma \xi \rangle,
\end{equation}
which shows that $\langle \Psi,\Delta_{U_t} \Psi\rangle \leq 0$ for
small values of $\varphi=2\, \mathrm{arc\, ctg} \,(k \pi/2\epsilon)$. Notice that
the normalization of the edge state $\Psi$
$$||\Psi||^2 = \int_\Omega (\Psi^\dagger,\Psi)_x \, d\mu_\eta (x) \geq
\frac{\pi(1 - \delta)}{2\epsilon} \, ||\xi||^2\,\int_0^{2\pi}\, ds\,
{\rm e}^{-2 k \tan s}$$
vanishes in the limit $t \to 0$ but it is always
a positive factor for $t \neq 0$ which preserves the bound given in Eq. \eqref{bbc}.  
Moreover, the nature of the edge state $\Psi$ also shows the existence
of a ground state $\Psi_0$ with negative energy which is an edge state.
The energy $E_0$ of this state goes to  $-\infty$ as $t \to 0$,
whereas the edge state $\Psi_0$ shrinks to the edge disappearing from
the spectrum of $\Delta_{U_t}$ in that limit.
\hfill$\Box$

\bigskip

Although the role of boundary conditions in the two Cayley submanifold
$\CC_\pm$ is quite similar from the mathematical point of view,
the boundary conditions are quite different from the physical
viewpoint. In particular, an analysis along the lines of the proof
of the above theorem leads to the same inequality as in Eq. \eqref{bbc} but with
$$k = \frac{2\epsilon}{\pi}\tan \frac{t}{2}$$
which points out the existence of  edge states with very large
(positive) energy as $t \to 0$. It can also be shown that
in that limit one energy level crosses the  zero energy level
becoming a zero mode of the Laplacian operator.
Therefore, the role of boundary conditions in $\CC_-$ (e.g. Dirichlet)
is very different of that of boundary conditions in $\CC_+$ (e.g. Neumann).

Notice that the result of the theorem does not require $U$ to be
in the self--adjoint Grassmannian $\CM'$. This is specially interesting,
because there is a very large family of boundary conditions which
do not belong to $\CM'$. In particular, boundary conditions implying
a topology change in higher dimensions
are not in  $\CM'$ because  the corresponding
unitary operators in $\CU(L^2(\Gamma, \mathbb{C}^r))$ present an infinity of  eigenvalues
$\pm 1$ which implies that $U$ cannot be of the form $\mathbb{I} + K$ with
$K$ Hilbert-Schmidt.
 Indeed, all boundary conditions which involve a change of topology, i.e.
gluing together domains $\CO_1$, $\CO_2$ of the boundary $\Gamma$, 
belong to $\CC_-\cap\, \CC_+$.
This property follows from the fact that the boundary conditions imply
that the boundary values $\varphi,\dot\varphi$ are related in the domains
that are being glued together, i.e. $\varphi(\CO_1)=\varphi(\CO_2),
\dot\varphi(\CO_1)=-\dot\varphi(\CO_2)$, respectively.   These 
requirements imply that the unitary operator $U$ corresponding
to this boundary condition is identically $U=\mathbb{I}$ on the subspace of
functions such that
$\varphi(\CO_1)=\varphi(\CO_2)$ and $U=-\mathbb{I}$ on the subspace of
functions such that $\varphi(\CO_1)=-\varphi(\CO_2)$. 
Since both subspaces
are infinite--dimensional for manifolds $\Omega$ of dimension larger than 1,
it is clear that those operators $U$ do not belong to $\CC'_-\cap\, \CC'_+$.
However the result of Theorem \ref{edge_states} implies that there always exists a
boundary condition close to one involving the gluing of
the domains with very large negative
energy levels. This means that Cayley manifold $\CC_-\cap\,\CC_+$ 
is very special and that topology change involves an interchange of an
infinite amount of quantum energy. These results might have relevant
implications in quantum gravity and string theory.



\section{Lecture 3:  Elliptic and self--adjoint extensions of Dirac operators}

\subsection{Dirac operators} 

As it was indicated in the introduction, Dirac operators constitute an important class of 
elliptic operators, to the extent that all relevant elliptic operators 
arising in Geometry and Physics are in one way or the other related to them.  
Let us set the ground to discuss them (see for instance \cite{La89} and \cite{Bo93}).  We will consider again a
Riemannian manifold $(\Omega, \eta )$ with smooth boundary
$\partial\Omega$.  We denote by $\mathrm{Cl}(\Omega)$ the Clifford
bundle over $\Omega$ defined as the algebra bundle whose fibre at $x\in
\Omega$ is the Clifford algebra $\mathrm{Cl}(T_x\Omega)$ generated by vectors $u$ in
$T_x\Omega$ satisfying ther relations
$$u\cdot v + v\cdot u = - 2\eta (u,v)_x ~~~~~~ \forall u,v\in T_x\Omega .$$ 

Let $\pi\colon S \to \Omega$ be a $\mathrm{Cl} (\Omega)$-complex vector bundle over $\Omega$, i.e.,
for each $x\in \Omega$, the fibre $S_x$ is a
$\mathrm{Cl}(\Omega)_x$-module, or in other words, there is a representation of
the algebra $\mathrm{Cl}(\Omega)_x$ on the complex space $S_x$ by complex automorphisms.  
We will represent with the same symbol the vector $u\in \mathrm{Cl} (\Omega)$ and
the automorphism of $S$ defined by $u$,  $\xi\mapsto u\cdot \xi$ for all
$\xi\in S$.  We will also call Clifford multiplication of
$\xi$ by $u$ the action of the automorphism defined
by the vector $u$ on the element $\xi$ of $S$.

We will assume in what follows that the bundle $S$ carries a hermitian metric denoted by
$(\cdot,\cdot )$ such that Clifford multiplication by unit vectors in $T\Omega$
is unitary:
\begin{equation}\label{unit_cl} ( u\cdot \xi, u\cdot \zeta )_x = ( \xi,\zeta )_x ,
\end{equation}
for all $\xi,\zeta \in S_x$, $u\in T_x\Omega$, $x\in \Omega$ and $||u||^2 = 1$.
Finally, we will assume that there is an Hermitean connection $\nabla$ on $S$ such
that 
\begin{equation}\label{derD} \nabla ( V \cdot \xi ) = (\nabla_\eta V)\cdot \xi +
V\cdot \nabla \xi ,
\end{equation}
where $V$ is a smooth section of the Clifford bundle $\mathrm{Cl} (\Omega)$, $\xi \in \Gamma
(S)$ and $\nabla_\eta$ denotes the canonical connection on
$\mathrm{Cl}(\Omega)$ induced by the Riemannian metric $\eta$ on $\Omega$.  A
bundle $S$ with the structure described above is commonly called a Dirac
bundle \cite{La89} and they provide the natural framework to define Dirac
operators.  Thus, if $\pi\colon S\to \Omega$ is a Dirac bundle, 
and we denote by $\Gamma^\infty (S)$ the space of smooth sections of the bundle map $\pi$, 
we can define a
canonical first-order differential operator $D\colon \Gamma^\infty (S) \to
\Gamma^\infty (S)$ by setting
$$ D\xi = e_j \cdot \nabla_{e_j} \xi ,$$
where $e_j$ is any orthonormal frame at $x\in \Omega$. 

There is a natural inner product $\langle \cdot, \cdot \rangle$ on $\Gamma (S)$ induced from the
pointwise inner product $(\cdot, \cdot )$ on $S$ by setting
$$ \langle \xi , \zeta \rangle = \int_\Omega  (\xi (x), \zeta (x))_x \vol_\eta (x) .$$ 
We will denote the corresponding norm by
$||\cdot ||_2$ and $L^2 (S)$ will denote the Hilbert space of square integrable sections 
of $S$.  


Giving a section $\xi$ which is square integrable, 
we will say that the 1-form $\beta$ in $\Omega$ with
values in $S$ is a weak covariant derivative of $\xi$
if it is square integrable and for every section $\zeta\in\Gamma^\infty_0(S)$, 
i.e., a smooth section of $S$ with compact support contained in the interior of $\Omega$,  we have:

\begin{equation}\label{weak}  \int_\Omega (\xi (x), \nabla_V \zeta (x) )_x \vol_\eta (x) = - 
\int_\Omega ( i_V\beta (x), \zeta (x) )_x \vol_\eta (x),
\end{equation}
for all vector fields $V$ in $\Omega$.
We consider the completion of
$\Gamma^\infty (S)$ with respect to the Sobolev norm $|| \cdot ||_{1,2}$ defined as:
\begin{equation}\label{sobov_k} 
||\xi ||_{k,2}^2 = \int_\Omega (\xi (x), (I +
\nabla^\dagger\nabla)^{k/2} \xi (x) )_x \vol_\eta (x) ,
\end{equation}
with $k =1$, where $\nabla^\dagger$ is the formal adjoint operator to $\nabla$ in $\Gamma_0^\infty(S)$. 
This Hilbert space will be denoted by $\mathcal{H}^1(S)$.

Moreover, it happens that the Dirac operator $D$ defined on $\mathcal{H}^1 (S)$ is not self--adjoint.
However it is immediate to check that the Dirac operator $D$
is symmetric in the space of smooth 
sections of $S$ with compact support contained in the interior of $\Omega$.  
In fact, after
integration by parts we obtain immediately,
\begin{equation}\label{symD} 
\langle D \xi , \zeta \rangle = \langle \xi, D \zeta
\rangle , \qquad \forall \xi, \zeta \in \Gamma^\infty_0 (S).
\end{equation} 
The operator $D$ with domain $\Gamma^\infty_0 (S)$
is closable on $L^2 (S)$ and its closure is the completion of $\Gamma_0^\infty
({S})$ with respect to the norm $||. ||_{1,2}$. 
Such domain will be denoted by $\mathcal{H}_0^1 (S)\subset \mathcal{H}^1(S)$.    
If we denote by $D_0$ the operator $D$ with domain
$\mathcal{H}_0^1 (S)$ then we are looking for
extensions of $D_0$ with domains 
dense subspaces of $\mathcal{H}^1 (S)$ containing $\mathcal{H}_0^1(S)$ and such
that the boundary terms obtained integrating by parts
will vanish.  Then the self--adjoint
extensions $D_s$ of $D_0$ we are looking for will be defined on subspaces $\dom (D_s)$ such
that 
$$\mathcal{H}_0^1 (S) = \dom (D_0) \subset \dom (D_s) = \dom (D_s^\dagger)
\subset \dom (D) = \mathcal{H}^1 (S), $$
and $D_s\xi = D \xi$ for any $\xi\in \dom (D_s)$.   

Our first aim will be to
characterize such subspaces using the geometry of some Hilbert
spaces defined on the boundary of $\Omega$.  
To achieve it, we will derive the expression of the boundary form obtained intregrating by parts.

Let $x\in \Omega$ and $e_j$ a local self--parallel orthonormal frame defined in a neighborhood of
$x$, $\nabla_{e_j} e_i = 0$ for all $i,j$.   It is easy to
see that such frame does always exists.   Then, if
$\xi$, $\zeta$ are sections of $S$, then they define a unique vector field
$X$ in a neighborhood of $x$ by the condition
\begin{equation}\label{auxiliary}
\eta ( X, Y ) = - ( \xi, Y\cdot \zeta ) ,
\end{equation}
for any vector field $Y$.  Then, we have that:
$$( D\xi, \zeta )_x = ( e_j \nabla_{e_j} \xi, \zeta
)_x = - \mathcal{L}_{e_j} ( \xi, e_j\zeta )_x + ( \xi , D\zeta )_x ,$$
but, 
$$\mathrm{div} (X) = \eta (\nabla_{e_j}X, e_j ) ,$$
hence using Eq. \eqref{auxiliary} again,
$$\mathrm{div} (X) = \mathcal{L}_{e_j} \eta_x ( X, e_j ) - \eta_x (X, \nabla_{e_j}e_j) =  - \mathcal{L}_{e_j} ( \xi , e_j\cdot \zeta )_x .$$ 
Namely,
$$ ( D\xi (x), \zeta (x) )_x - (\xi (x) , D\zeta (x))_x = \mathrm{div} _x
(X) .$$ 
Integrating the previous equation we find,
$$ \langle D\xi, \zeta\rangle - \langle \xi , D\zeta \rangle = \int_\Omega
\mathrm{div} _x (X) \vol_\eta (x) = \int_\Omega  (i_X d \vol_\eta) = \int_{\partial
\Omega} i^* (i_X \vol_\eta) ,$$
where we denote by $i\colon \partial
\Omega \to \Omega$ the canonical inclusion. 
If $\nu$ denotes
the inward unit vector on the normal bundle to
$\partial \Omega$, the volume form $\vol_\eta$ can be written on a neighborhood of $\partial \Omega$ as
$\theta\wedge \vol_{\partial\eta}$, where $\vol_{\partial\eta}$ is an
extension of the volume form defined on
$\partial \Omega$ by the restriction of $\eta$, and $\theta$ is the
1-form such that $\theta (Y) = \eta (Y, \nu)$ for all $Y$.   Then  we get,
$$ i_X \vol_\eta = (i_X\theta) \vol_{\partial\eta} = \eta (X,\nu )
\vol_{\partial \eta} = ( \xi, \nu\cdot \zeta ) \vol_{\partial\eta} .$$
Thus, finally, we obtain:
\begin{equation}\label{int_part_D1} 
\langle D\xi, \zeta\rangle - \langle \xi ,
D\zeta \rangle =  \int_{\partial \Omega} i^* (\nu\cdot \xi,
\zeta ) \vol_{\partial \eta}(x) .
\end{equation}
We have obtained in this way the
Lagrange's boundary bilinear form
\Eq{\label{int_part_D2} \Sigma (\xi , \zeta ) = \int_{\partial \Omega}
( \nu(x)\cdot \xi (x), \zeta (x) )_x \vol_{\partial \eta} (x) ,}
responsible for the non self--adjointeness of the Dirac operator $D$ in
$\mathcal{H}^1 (S)$. 

\subsection{The geometric structure of the space of boundary data}

We will denote by $\partial S$ the restriction of the Dirac bundle $S$
to $\Gamma = \partial\Omega$, i.e., $\partial S = S\mid_{\Gamma}$ which is a 
bundle over $\Gamma$, $\partial \pi \colon \partial S \to \Gamma$ with $\partial \pi = \pi \mid_{\partial S}$.  It is noticeable that $\partial S$ becomes a Dirac bundle over $\Gamma = \partial
\Omega$ with the inner product $\langle \cdot, \cdot  \rangle_{\partial \Omega}$ induced from the Hermitean
product on $S$ by restricting it to $\Gamma$ and the induced
Hermitean connection $\nabla_{\partial\Omega}$, defined again by restricting the
connection $\nabla$ on $S$ to sections along $\partial\Omega$.  Thus the
boundary Dirac bundle $\partial S$ carries a canonical
Dirac operator denoted by $D_{\partial \Omega}$ and called the tangential Dirac operator.  Notice that $\Gamma = \partial \Omega$ is a manifold without boundary, thus the boundary Dirac operator
is essentially self--adjoint and possesses a unique self--adjoint extension
(see for instance \cite{La89}, Thm. 5.7; this fact will also follow from our main
theorem in this section).  

We will denote as before by $L^2 (\partial S)$ the Hilbert space of
square integrable sections of $\partial S$ and by $\langle
\cdot ,\cdot \rangle_{\partial\Omega}$ its Hilbert product structure
\begin{equation}\label{bounD} 
\langle \phi , \psi \rangle_{\partial \Omega} = \int_{\partial
\Omega} (\phi(x), \psi (x) )_x \vol_{\partial\eta} (x) .
\end{equation}
Because of the trace theorem the restriction map 
$i^* \colon \Gamma^\infty (S) \to \Gamma^\infty (\partial S)$, 
$\xi \mapsto \phi :=i^* \xi$, extends to a continuous linear map, called the trace or boundary map again:
\begin{equation}\label{trace_dirac}
b\colon \mathcal{H}^1(S) \to \mathcal{H}^{1/2}(\partial S) \subset L^2 (\partial S)  .
\end{equation}
Moreover $b$ induces a homeomorphism $\tilde{b} \colon \mathcal{H}^1(S) /\ker b \to \mathcal{H}^{1/2}(\partial S)$, with $\ker b = \mathcal{H}_0^1(S)$ (see for instance \cite{Ad75}, Thm. 7.53).    Hence, if $\dom (D_s) \subset \mathcal{H}^1(S)$ is the domain of a self--adjoint extenstion of the Dirac operator $D$, then $b(\dom (D_s)) \subset \mathcal{H}^{1/2}(\partial S)$ is a closed subspace. 
The Hilbert space $L^2 (\partial S)$ will be called the Hilbert space of
boundary data for the Dirac operator $D$ and will be denoted in what follows by
$\mathcal{H}_D$.  It carries an important extra geometrical structure induced by the
Lagrange boundary form $\Sigma$,  Eq. (\ref{bounD}).   The normal vector field along $\Gamma$, defines a smooth section $\nu$ of the Clifford bundle $\mathrm{Cl}(\partial \Omega)$ over
$\Gamma = \partial \Omega$, thus
$\nu$ defines an automorphism of the Dirac bundle $\partial S$
over $\partial \Omega$, 
\begin{equation}\label{section_nu}
\nu\colon \Gamma^\infty (\partial S) \to \Gamma^\infty (\partial S) ,\quad \nu (\phi ) (x) = \nu (x) \cdot \phi (x), \quad \forall x\in \Gamma,\quad  \phi\in \Gamma^\infty (\partial S) .
\end{equation}
Such automorphism extends to a
continuous complex linear operator of $\mathcal{H}_D$ denoted now by $J_D$.  
Because $\nu^2 = - 1$ in the Clifford algebra, such operator $J_D$ verifies
$J_D^2 = -\mathbb{I}$.  In addition, because of the Dirac bundle structure, Eq.
(\ref{unit_cl}), $J_D$ is also an isometry of the Hilbert space product, this is:
$$ \langle J_D\phi, J_D\psi \rangle_{\partial\Omega} =  \langle \phi,\psi
\rangle_{\partial\Omega}, ~~~~~ \quad  \forall \phi, \psi \in \mathcal{H}_D ,$$ 
i.e., $J_D$ defines a compatible complex structure on
$\mathcal{H}_D$.  

More generally, given a complex Hilbert space
$\mathcal{H}$ with inner product $\langle \cdot, \cdot \rangle$ and a compatible
complex structure $J$ we can define a new continuous bilinear form $\omega$ by
setting,
$$ \omega (\varphi,\psi) = \langle J\varphi, \psi\rangle , ~~~~\quad \forall
\varphi,\psi \in \mathcal{H} .$$
Such structure is skew-Hermitiean in the sense that 
$$\omega (\varphi , \psi ) = - \overline{\omega (\psi, \varphi) } .$$
If the Hilbert space $\mathcal{H}$ would be real $\omega$ will define a symplectic structure on $\mathcal{H}$.  
In any case the real part of $\omega$ will always define a real symplectic
structure on $\mathcal{H}_D$ viewed as a real space, very much as
the imaginary part of a Hermitean structure on a complex Hilbert 
space defines a symplectic structure on its realification.
We call the space $\mathcal{H}$ with the Hermitean
and skew-Hermitean structures $\langle \cdot, \cdot \rangle$ and $\omega$, a symplectic--Hermitean
linear space.

Any symplectic--Hermitean linear space carries a natural polarization.
In fact, the compatible complex structure allows to decompose the Hilbert space
$\mathcal{H}$ as $\mathcal{H}_+ \oplus \mathcal{H}_-$ where $\mathcal{H}_\pm$ 
are the closed
eigenspaces of $J$ of eigenvalues $\mp i$, that is $\phi_\pm \in \mathcal{H}_\pm$ if $J\phi_\pm = \mp
i \phi_\pm$.  The subspaces $\mathcal{H}_\pm$ are orthogonal because:
$$
\langle \phi_+ , \phi_- \rangle = \langle J\phi_+ , J\phi_- \rangle = 
\langle i\phi_+ , -i\phi_- \rangle = - \langle \phi_+ , \phi_- \rangle .
$$

Notice that 
the Hilbert space $\mathcal{H}_D$ carries already another complex structure, 
denoted by $J_0$, which is simply multiplication by $i$.  Both complex structures are 
compatible in the sense that $[J_D,J_0] = 0$    
because the Dirac bundle $S$ is a $\Cl(\Omega)$--complex bundle.   

Hence, the previous discussion shows that the Hilbert space of boundary data $\mathcal{H}_D$ for the Dirac operator $D$ is a polarized Hilbert space carrying a compatible complex
structure $J_D$ and the corresponding skew--Hermitean structure denoted in what follows
by  $\omega_D$.  
Using these structures the Lagrange boundary form $\Sigma$ is written as:
\begin{equation}\label{bound_D}
\Sigma (\xi, \zeta)= \omega_D (b(\xi), b(\zeta)) ,
\qquad \forall \xi, \zeta\in \mathcal{H}^1 (S) .
\end{equation}
From Eq. \eqref{bound_D} we see immediately that symmetric
extensions $D_s$ of $D$ will be defined in domains $\dom (D_s)$ such that
their boundary image $b(\dom (D_s))$ are isotropic subspaces $W$ of
$\omega_D$, i.e., such that the r.h.s. of Eq. \eqref{bound_D} vanishes for all $b(\zeta),b(\xi) \in W$.    Moreover if the extension $D_s$ is
self--adjoint, such domains must verify that $b(\dom (D_s)) = b(\dom
(D_s^\dagger))$ thus, they must be maximal subspaces with this property. 
We have thus proved the first part of the following theorem:

\begin{theorem}\label{unit_D}  Let $(\Omega, \eta)$ be a spin manifold 
with smooth boundary $\partial \Omega$, $\pi\colon S \to \Omega$ a Dirac bundle and $D$ a Dirac operator on $S$.  Then the symplectic--Hermitean boundary data Hilbert space $(\mathcal{H}_D, J_D, \omega_D)$ carries a natural polarization
$\mathcal{H}_D = \mathcal{H}_+ \oplus \mathcal{H}_-$ and self--adjoint extensions of the Dirac
operator $D$ with domain in $\mathcal{H}^1(S)$ are in one--to--one correspondence with subspaces of the boundary Hilbert space $\mathcal{H}_D$ which are maximally $\omega_D$--isotropic closed subspaces on $\mathcal{H}^{1/2}(\partial S)$. 
The domain of anyone of these extensions is the inverse image by the
boundary map $b$ of the corresponding isotropic subspace.   Moreover, 
each maximally closed $\omega_D$--isotropic subspace $W$ of $\mathcal{H}_D$
defines an unitary operator $U\colon \mathcal{H}_+\to \mathcal{H}_-$ and conversely.   
\end{theorem}

{\parindent 0cm \emph{Proof:}} 
Let $W$ be a closed $\omega_D$-isotropic subspace of $\mathcal{H}_D$. 
Then, $b^{-1} (W)$ is a closed subspace of $\mathcal{H}^1 (S)$ containing $\mathcal{H}_0^1
(S)$.  Let $D_W$ be the extension of $D$ defined on $b^{-1} (W)$ and
compute $D_W^\dagger$.  If $b(\xi),b(\zeta) \in W$, then $\langle
D_W^\dagger\xi, \zeta\rangle = \langle \xi, D_W \zeta \rangle + \omega
(b(\xi), b(\zeta)) = \langle \xi, D_W \zeta \rangle$ because $W$ is
$\omega_D$ isotropic.  This shows that $b^{-1}(W) \subset \dom
(D_W^\dagger)$. If there were $\xi\in \dom (D_W^\dagger) - b^{-1}(W)$,
then, the same computation shows that $\omega_D (b(\xi), \phi) = 0$
for all $\phi\in W$, and the subspace $W^\prime = W \oplus \langle
b(\xi) \rangle$ will be $\omega_D$-isotropic, which is contradictory.  Thus
$\dom (D_W) = \dom (D_W^\dagger)$ and the extension is self--adjoint.  
The converse is proved similarly.

Let us consider now a closed maximal $\omega_D$-isotropic subspace $W$.
Let us show that $W$ is transverse to $\mathcal{H}_\pm$.  Let $\phi\in W\cap
\mathcal{H}_\pm$, then $0= \omega_D (\phi, \phi) = \langle J\phi, \phi \rangle
= \mp i || \phi ||^2$, then $\phi = 0$.  Then, the subspace $W$ defines
the graph of a continuous linear operator $U\colon \mathcal{H}_+ \to \mathcal{H}_-$ 
and vectors $\phi = \phi_+ + \phi_- \in W$ have the form $\phi_- = U\phi_+$.  Then, the
$\omega_D$-isotropy of $W$ implies,
\begin{eqnarray*}
 0 = \omega_D (\phi_+ + U\phi_+, \psi_+ + U\psi_+) &=& \langle i\phi_+
- iU\phi_+ , \psi_+ + U\psi_+ \rangle \\ &=& -i \langle \phi_+, \psi_+
\rangle + i\langle U\phi_+ , U\psi_+ \rangle ,
\end{eqnarray*}
for every $\phi_+, \psi_-\in \mathcal{H}_+$,
that proves that $U$ is an isometry.  
\hfill$\Box$

\bigskip

We can use anyone of these extensions, say one defined by
a subspace $W_0$, to identify the space of self--adjoint extensions
of $D$ with the group of unitary transformations of $\mathcal{H}_+$.   
Thus if $W$ is a closed maximally $\omega_D$-isotropic subspace,
let $\phi_W\colon \mathcal{H}_+ \to \mathcal{H}_-$ be the isometry defined
by it because of Thm. \ref{unit_D}, then we associate to it the 
map $\phi = \phi_0^{-1} \circ \phi_D$ where $\phi_0$ is the isometry
associated to  $W_0$.  It is trivial then to check that $\phi$ is indeed
a unitary operator on $\mathcal{H}_+$.  

From the previous discussion we can conclude that the
space of self--adjoint extensions of the Dirac operator 
$D$ can be naturally identified with $U(n)$ if $\dim\,\, \Omega = 1$, $n$ is the number of connected components of 
$\partial \Omega$, however it is contractible if $\dim\,\, \Omega > 1$.

\subsection{The Cayley transformation at the boundary}\label{Cayley_D}

In spite of the inherent interest of the results described in
the previous section, sometimes (as in the case of Laplace operators)
it is more useful to have an alternative description of self--adjoint extensions in 
terms of self--adjoint operators at the boundary.  For that we
will use the Cayley transformation again in a new fashion.  

For that purpose, we define the space $\mathcal{H}_J$ as the
graph of the operator $J_D \colon \mathcal{H}_D \to \mathcal{H}_D$,
i.e., 
$$ \mathcal{H}_J = \set{ (\xi, J_D\xi )\in \mathcal{H}_D \times \mathcal{H}_D \mid \xi \in
\mathcal{H}_D} .$$
The natural projection restricted to $\mathcal{H}_J$ defines an isometry among
$\mathcal{H}_D$ and the later.   
The space $\mathcal{H}_J$ carries a natural polarization induced from 
the one on $\mathcal{H}_D$.  Thus we define the closed orthogonal subspaces
$$\mathcal{L}_\pm = \set{ (\xi_\pm, J\xi_\pm )\in \mathcal{H}_D \times \mathcal{H}_D \mid
\xi_\pm \in \mathcal{H}_\pm} ,$$
and clearly,
$$ \mathcal{H}_J = \mathcal{L}_+ \oplus \mathcal{L}_-, \quad \mathcal{L}_+^\perp =
\mathcal{L}_-  .$$

We define the Cayley transformation on the
polarized boundary Hilbert space 
$\mathcal{H}_D = \mathcal{H}_+ \oplus \mathcal{H}_-$ as the
continuous isomorphism
$C \colon \mathcal{H}_D = \mathcal{H}_+ \oplus \mathcal{H}_- \to \mathcal{H}_J =
\mathcal{L}_+ \oplus \mathcal{L}_-$ 
defined by 
\begin{equation}\label{cayley_dirac} 
C(\xi_+, \xi_-) = \left( \mitad (\xi_+ + \xi_-), -\frac{i}{2}(\xi_+
-  \xi_-) \right) ,
\end{equation} 
for every $\xi_\pm \in \mathcal{H}_\pm$.   We will also use the notation:
\begin{equation}\label{C+-}
C_+ (\xi_+, \xi_-) =  \mitad (\xi_+ + \xi_-) \in \mathcal{L}_+, \quad C_- (\xi_+, \xi_-) =  \frac{i}{2} (\xi_+ - \xi_-) \in \mathcal{L}_-,
\end{equation}
thus $C = (C_+,C_-)$.
If we denote by $\psi_+ = \mitad (\xi_+ + \xi_-)$ and $\psi_- = -\frac{i}{2}(\xi_+
-  \xi_-) $, we have that $J_D(\psi_+) = \psi_-$ as it should be.
The map $C$, which is $J_0$ complex, transforms
the complex structure $J_D$ into
$J = C J_D C^{-1} (\phi_+, \phi_-) = (-\phi_-, \phi_+)$ and the
symplectic-hermitian structure $\omega_D$ is transformed into the
bilinear form
\begin{equation}\label{omega_C} \sigma (\phi_+,\phi_+; \psi_+, \psi_-) = 2i (\langle
\phi_+, \psi_- \rangle - \langle \phi_- , \psi_+ \rangle ) .
\end{equation}
    
Let $U$ be an isometry $U\colon \mathcal{H}_+ \to \mathcal{H}_-$.   
Then we have that the elements of
$\mathcal{H}_D$ in the graph of $U$ verify $\xi_- = U\xi_+$.  Using the
Cayley transformation Eq. \eqref{cayley_dirac} we will obtain that, 
$\psi_+ = \mitad (\xi_+ + U\xi_+)$, and  $\psi_- = -\frac{i}{2}(\xi_+
-  U\xi_+) $ hence, $(I + U)\psi_- = -i(I-U)\psi_+$.  
Thus we conclude that the graph of the isometry $U$ is mapped
into the subspace $W_U$ of $\mathcal{H}_J$ defined by 
$$ W_U = \{(\psi_+, \psi_-)\in \mathcal{H}_J \mid  (I + U)\psi_- = -i(I-U)\psi_+\} .$$

Let $\mathcal{H} = \mathcal{K}_+ \oplus \mathcal{K}_-$ be a polarized Hilbert space.
Let $W$ be a subspace of $\mathcal{H}$, the adjoint of
$W$ is the subspace denoted by $W^\dagger$ and defined by:
$$ 
W^\dagger = \set{(\psi_+, \psi_-)\in \mathcal{K}_+\oplus \mathcal{K}_- \mid
\langle \phi_+, \psi_- \rangle = \langle \phi_-, \psi_+ \rangle, ~~
\forall (\phi_+, \phi_-)\in W} .
$$
The subspace $W$ is said to be symmetric if
$W\subset W^\dagger$ and self--adjoint if $W = W^\dagger$. 
Notice that an operator $A\colon \mathcal{K}_+ \to \mathcal{K}_-$ is self--adjoint 
if its graph is a self--adjoint subspace of the polarized Hilbert space 
$\mathcal{H} = \mathcal{K}_+ \oplus \mathcal{K}_-$. 

Now a simple computation shows that the subspace $W_U$ constructed previously
is a self--adjoint subspace of the polarized Hilbert space 
$\mathcal{H}_J = \mathcal{L}_+ \oplus \mathcal{L}_-$.
The subspace $W_U$ is transverse to $\mathcal{L}_+$, i.e.,
$W_U \cap \mathcal{L}_+ =0$, then it is the graph of a
self--adjoint operator $A_U\colon \mathcal{L}_+ \to \mathcal{L}_-$.  
In this sense, the Cayley transformation operator $A_U$ of 
any isometry $U$ is self--adjoint.   
Moreover, it is clear that self--adjoint subspaces are maximally isotropic subspaces of the bilinear
form $\sigma_D$ given by Eq. (\ref{omega_C}).  But $\sigma_D$ is the
transformed bilinear form on $\mathcal{H}_D$ by the Cayley transformation, then,
maximally $\sigma_D$-isotropic subspaces correspond to maximally
$\omega_D$-isotropic subspaces, in other words, the Cayley transformation is a one-to-one map
among isometries $U\colon \mathcal{H}_+ \to \mathcal{H}_-$ and
self--adjoint operators $A \colon \mathcal{L}_+ \to \mathcal{L}_-$.

We will consider the topology on
the spaces of isometries $U\colon \mathcal{H}_+ \to \mathcal{H}_-$
and self--adjoint operators $A \colon \mathcal{L}_+ \to \mathcal{L}_-$ induced by the norm
operator topology.  We can summarize the previous discussion in
the following theorem.

\begin{theorem}  The Cayley transform $C\colon \CH_D \to \CH_J$ defined by Eq. \eqref{cayley_dirac} defines a homeomorphism between the 
space of isometries $\CU(\mathcal{H}_+, \mathcal{H}_-)$ from 
$\mathcal{H}_+$ to $\mathcal{H}_-$ and the space of self--adjoint operators
$\mathcal{S}(\mathcal{L}_+ ,\mathcal{L}_-)$. 
Moreover, the self--adjoint extensions of the Dirac operator 
$D$ are in one-to-one correspondence
with the self--adjoint operators $\mathcal{S}(\mathcal{L}_+ ,\mathcal{L}_-)$. 
\end{theorem}

\subsection{The space of self--adjoint elliptic boundary conditions: the elliptic grasmannian}

In the previous section we have characterized self--adjoint extensions
of Dirac operators in terms of boundary data and we have seen that they 
can be globally described as the manifold of self--adjoint subspaces $W$ of the Hilbert space $\mathcal{H}_J$.  
However we have not considered yet along this discussion 
if the extensions $D_W$ of the Dirac operator $D$ obtained in this way 
are elliptic operators or not, i.e., if the
boundary data given by $W$ determines an elliptic boundary problem for
$D$ \cite{Ag65}, \cite{Re85}.  This is a crucial issue for applications 
of the theory because if the
extensions  considered are not elliptic the resulting operator could have, for instance, an
infinite number of zero modes, 
i.e., its kernel will be infinite dimensional, 
which will make it unsuitable for physical
applications.   Looking for elliptic extensions of the operator
$D$ is thus a natural demand both mathematically 
and regarding the eventual applications of them.  

As it was mentioned before the theory of elliptic boundary problems 
for Dirac operators was developed
in the seminal series of papers by Atiyah, Patodi and Singer \cite{At71}.  The
boundary conditions introduced there to study the index theorem for
Dirac operators in even-dimensional spin manifolds with boundary are
nowadays called Atiyah-Patodi-Singer (APS) boundary conditions.    
The crucial observation there was
that global boundary conditions were needed in order to obtain an elliptic
problem and this was completely different to the situation for second order 
differential operators
where for instance ``local'' Dirichlet conditions are elliptic.   
Later on such boundary
conditions were extended to include also odd dimensional spin manifolds
with boundary (see \cite{Da94} and references therein).  
More recently, E. Witten (\cite{Wi88}, Section II), pointed out the link between
elliptic boundary conditions for the Dirac operator on 2 dimensions and 
the infinite dimensional Grassmannian manifold.  The infinite
dimensional Grassmannian was introduced previously in the analysis of
integrable hierarchies and discussed extensively by
Segal and Wilson (see \cite{Se85}, \cite{Pr86} and references therein).  
Finally Schwarz and Friedlander \cite{Fr97} have presented a way to
extend Witten's analysis to arbitrary elliptic operators on arbitrary
dimensional manifolds with boundary.  The particular analysis for Dirac
operators follows from \cite{At71} but we want to point it
out here that it can be extended also to higher order operators.  More
comments on this will be found later on. 

The basic idea behind is that the space of zero modes of a Dirac operator $D$, 
$$\ker D = \set{\xi\in \Gamma^\infty (S) \mid D\xi = 0} ,$$
induces a subspace in the boundary $b(\ker D)$ that in general will be
infinite dimensional.   The way to restore ellipticity will be to
project down into a subspace such that the kernel and cokernel of the
operator in this subspace will be finite dimensional.    We shall perform
such analysis for Dirac operators (see \cite{Bo93} for a detailed discussion).

The analysis of such projection requires the description of
solutions near the boundary.   We can decompose the operator $D$ in a collar
neighborhood  $(-1,0]\times \partial \Omega$ of the boundary as 
$$D = \nu\cdot (\nabla_\nu + D_{\partial\Omega}) ,$$
where $D_{\partial\Omega}$ is the Dirac operator on the boundary bundle
$\partial S$.  We easily see that $b(\ker D)$ is spanned by the
eigenfunctions of $D_{\partial \Omega}$ with nonnegative eigenvalues
\cite{At71}.   Next Lemma will describe this subspace in terms of the
boundary operator $D_{\partial\Omega}$ and subspaces of $\mathcal{H}_D$
or better $\mathcal{H}_J$.   In fact we get:

\begin{lemma} Let $\Omega$ be a compact manifold with smooth boundary $\partial \Omega$, then with the notations above, we have:
$$ C(b(\ker D )) = \mathcal{L}_+ .$$
\end{lemma}

Proof.  Because $\partial \Omega$ is a closed manifold $D_{\partial\Omega}$ is an
essentially self--adjoint elliptic differential operator.   Moreover the following computation shows
that $D_{\partial\Omega}$ anticommutes with $J_D$.
Namely,
$$J_D D_{\partial\Omega} = D - J_D\nabla_\nu ,$$
but it is easy to check that:
$$ J_D D = - D\nu - 2 \nabla_\nu ,$$
hence
$$ J_D D J_D = D - 2\nabla_\nu J_D .$$
Moreover,
$$ D_{\partial \Omega} J_D = - J_DD J_D - \nabla_\nu J_D ,$$
and, finally:
\begin{eqnarray*} J_D D_{\partial\Omega} + D_{\partial\Omega} J_D & = & D -  J_D \nabla_\nu - J_D D J_D  -
\nabla_\nu J_D = \\ & = & D -  J_D \nabla_\nu  - D + 2\nabla_\nu J_D - J_D \nabla_\nu =   0 ,
\end{eqnarray*}
hence
$D_{\partial\Omega} J_D = - J_D D_{\partial \Omega}$.  

The Dirac Laplacian $D_{\partial\Omega}^2$ is a non-negative essentially self--adjoint elliptic operator
with a real discrete spectrum $\mathrm{Spec} (D_{\partial\Omega}^2) = \set{\lambda_k
\mid 0 = \lambda_0 < \lambda_1 < \cdots }$ with finite dimensional eigenspaces 
$$E(\lambda_k) = \set{\phi_k \in \mathcal{H}_D \mid D_{\partial\Omega}^2\phi_k = \lambda_k
\phi_k}.$$  
The kernel $K$ of $D_{\partial\Omega}$ agrees with $\ker
D_{\partial\Omega}^2$ and with $E(0)$.  We have thus the following
orthogonal decomposition of $\mathcal{H}_D$,
$$ \mathcal{H}_D = \bigoplus_{k=0}^\infty E(\lambda_k) = K \oplus
\left(\bigoplus_{k=1}^\infty E(\lambda_k)\right) .$$
On the other hand the polarization $\mathcal{H}_D = \mathcal{H}_+ \oplus \mathcal{H}_-$
defined by the compatible complex structure $J_D$, $J_D \mid_{\mathcal{H}_\pm } =
\mp i \mathrm{I}$, induces a decomposition of the eigenspaces
$E(\lambda_k)$ as
$$ E(\lambda_k ) = E_+ (\lambda_k) \oplus E_- (\lambda_k) ;   \quad E_\pm (\lambda_k)= E(\lambda_k) \cap \mathcal{H}_\pm .$$

Moreover, $D_{\partial\Omega}$ restricts to a map $D_k =
D_{\partial\Omega} Ê\mid_{E(\lambda_k)} \colon E(\lambda_k) \to
E(\lambda_k)$ and because anticommutes with $J_D$, we have that $D_k
\colon E_\pm (\lambda_k) \to E_\mp (\lambda_k)$, thus $D_k$ has the block structure,

$$ D_k = \left[\begin{array}{c|c} 0 & D_k^+ \\ \hline D_k^- & 0\end{array}\right] .$$
In addititon, because $D_k$ is self--adjoint, $(D_k^-)^\dagger = D_k^+$.  On the
other hand $D_k^2 = D_{\partial\Omega}^2\mid_{E(\lambda_k)} = \lambda_k
I$, hence the spectrum of $D_k$ on $E(\lambda_k)$ is
$\pm\sqrt{\lambda_k}$.   The operator $D_k$ is invertible in
$E(\lambda_k)$ for $k\geq 1$, hence $\dim\,  E_+ (\lambda_k) = \dim \, E_-
(\lambda_k)$.  Moreover $K = K_+ \oplus K_-$, and $\dim\,  K_+ = \dim \, 
K_-$.   Notice that the index of the operator $D_0^+$ is zero because of the cobordant invariance of the index and the fact that
$\partial \Omega$ is cobordant to $\emptyset$.   Thus we can choose an orthonormal basis
$\phi_{k,\alpha}^\pm \in E_\pm (\lambda_k)$, $\alpha = 1,\ldots, \dim \, 
E_\pm (\lambda_k)$,  such that
$$ D_k \phi_{k,\alpha}^\pm = \pm \sqrt{\lambda_k} \phi_{k,\alpha}^\mp
.$$
The Cayley transformation discussed in Section \ref{Cayley_D} diagonalizes the operators
$D_k$, and if we denote  by $\psi_{k,\alpha}^\pm = C_\pm (\phi_{k,\alpha}^+ , \phi_{k,\alpha}^-)$ it is clear that  Then, it is clear that $b(\ker D) = \mathcal{L}_+$ because $\mathcal{H}_D$ is spanned by nonnegative eigenspaces of $D_{\partial \Omega}$ and eq. (\ref{C+-}). \hfill$\Box$

\bigskip

Then we can conclude the discussion by stating the following proposition:

\begin{theorem}
Elliptic boundary conditions for the Dirac operator $D$ are in one-to-one
correspondence with the set of subspaces 
$W\subset \mathcal{H}_J$ such that $W\cap \mathcal{L}_+$ is finite dimensional.
\end{theorem}

Proof.  We want to characterize subspaces $W$ such that the solutions of the equation $D\xi = 0$
with boundary values on $W$ will be finite dimensional.  The orthogonal
projectors $\pr_\pm \colon\mathcal{H}_J \to \mathcal{L}_\pm$ are pseudodifferential
operators whose complete symbol depends only on the coefficients of $D$. 
Hence, because of the previous Lemma, $C\circ b(\ker D) = \mathcal{L}_+$, and elliptic
boundary conditions will be defined by subspaces $W\subset
\mathcal{H}_D$ such that $W\cap \mathcal{L}_+$ will be finite dimensional.
\hfill$\Box$

\bigskip

Notice that for the elliptic extensions of the Dirac operator determined by
the subspace $W$, the projection
$\pr_+\mid_W$ will have a finite dimensional kernel.  

Moreover $\mathrm{coker}\, D_W = \ker D_W^\dagger$ because
$$ \langle \zeta, D_W\xi \rangle =  \langle D_W^\dagger\zeta, \xi \rangle ,$$
hence, $\zeta \in \ker D_W^\perp$ iff $\zeta \in \ker D_W^\dagger$, and 
$\mathrm{coker}\, D_W$ is identified naturally with $\ker D_W^\perp$.
Hence, the cokernel of $\pr_+$ will have to be finite-dimensional if
$D_W^\dagger$ is elliptic too.  

Finally, if the extension
$D_W$ is elliptic, then there will exists left and right parametrics for it (see for
instance \cite{La89}), and
this will imply that the projection $\pr_-\mid_W$ will have to be a
compact operator.   Then we conclude from the previous discussion:

\begin{theorem}
The set of elliptic extensions of the Dirac operator $D$ is in one-to-one correspondence
with the points of the compact Grassmannian $\mathrm{Gr}_\mathcal{K}$, where
$\mathrm{Gr}_\mathcal{K}$ is defined as the set of closed subspaces $W$ of $\mathcal{H}_J$ such
that the projections $\pr_+ \colon W \to \mathcal{L}_+$ is a Fredholm operator and 
$\pr_- \colon W \to \mathcal{L}_-$ is a compact operator.
\end{theorem}

The set $\mathcal{K}$ of compact operators contains a 
distinguished subset, the Hilbert-Schmidt operators.
For technical reasons it is convenient to consider
a restriction of the infinite dimensional Grassmannian
to consider only those subspaces such that the 
projection on $\mathcal{L}_-$ is Hilbert--Schmidt.

We will say that closed subspaces $W$ of $\mathcal{H}_J$ satisfying that
the projection on the first factor $\pr_+\mid_W \colon W \to \mathcal{L}_-$ is a Fredholm operator
and the projection on the second factor $\pr_-\mid_W \colon W \to \mathcal{L}_+$ is Hilbert-Schmidt
define restricted elliptic extensions of the Dirac operator $D$.   Such
space will be called the elliptic infinite--dimensional Grassmannian of
$D$, or elliptic Grassmannian for short, and will be denoted by ${\mathrm{Gr}}'$ (compare with the definition of the self--adjoint Grassmannian for the Bochner Laplacian in Section \ref{selfadjoint_grassmannian}).   

The elliptic Grassmannian can be constructed also in terms of the
polarization $\mathcal{H}_+ \oplus \mathcal{H}_-$ 
instead of $\mathcal{L}_+ \oplus \mathcal{L}_-$.   
This is the approach taken for instance in \cite{Da94}.  
In such case, we relate self--adjoint
extensions of $D$ with isometries $U\colon \mathcal{H}_+ \to \mathcal{H}_-$, 
hence elliptic
boundary conditions correspond to isometries $U$ such that the projection
from its graph to $\mathcal{H}_+$ would be Fredholm and the projection onto
$\mathcal{H}_-$ would be Hilbert-Schmidt.  It is obvious that the Cayley
transformation  $C$ defines a one-to-one map from the
Grassmannian $\mathrm{Gr}(\mathcal{H}_-,\mathcal{H}_+)$ into
$\mathrm{Gr} (\mathcal{L}_-, \mathcal{L}_+)$ (the map is actually a diffeomorphism, see
below), but it is important to keep in mind that the objects in the two
realizations of the Grassmannian are different.  
In what follows we will omit the subindex to the different Hilbert
spaces $\mathcal{H}_D$ and $\mathcal{H}_J$ and they will be identified
by means of the Cayley transformation as indicated above.

We will call in what follows the elliptic boundary conditions defined by points in the
elliptic Grassmannian, generalized APS boundary conditions. 
The elliptic infinite--dimensional Grassmannian has an important
geometrical and topological structure.  We must recall first (see for
instance Pressley and Segal \cite{Pr86} for more details) that $\mathrm{Gr}'$ is
a smooth manifold whose tangent space at the point $W$ is given by
the Hilbert space of Hilbert--Schmidt operators 
$\mathcal{J}_2 (\mathcal{L}_-,\mathcal{L}_+)$, 
from $\mathcal{L}_-$ to $\mathcal{L}_+$.  
The group of linear continuous invertible operators $GL
(\mathcal{H})$ does not act on ${\mathrm{Gr}}'$ 
but only a subgroup of it, the restricted general
linear group $GL_{\res}(\mathcal{H})$, which defines the restricted unitary group
$U_{\res}(\mathcal{H}) = GL_{\res}(\mathcal{H})Ê\cap U(\mathcal{H})$.   The groups
$GL(\mathcal{H})$ and $U(\mathcal{H})$ are contractible but
$GL_{\res}(\mathcal{H})$ and $U_{\res}(\mathcal{H})$ are not.  The manifold ${\mathrm{Gr}}'$ is
not connected and is decomposed in its connected components defined by
the virtual dimension of their points which is simply the index of the
Fredholm operator $\pr_+\mid_W$, then, ${\mathrm{Gr}}' = \cup_{k\in \Z} \mathrm{Gr}^{(k)}$. 

The Grassmannian $\mathrm{Gr}'$ carries a natural K\"ahler
structure defined by the hermitian structure given by
$$ h_W (\dot A, \dot B ) = \Tr  (\dot A^\dagger \dot B ) ,$$
where $\dot A, \dot B\in T_W \mathrm{Gr}'$ are Hilbert-Schmidt
operators from $\mathcal{L}_-$ to $\mathcal{L}_+$.  The imaginary part defines a
canonical symplectic structure $\Omega$,
\begin{equation}\label{symp} 
\Omega_W (\dot A, \dot B) = -\frac{i}{2} \Tr (\dot
A^\dagger \dot B - \dot B^\dagger \dot A ) .
\end{equation}
The Grassmannian $\mathrm{Gr}'$ is cuasicompact in the sense that the only holomorphic
functions are constant.

\subsection{The self--adjoint Grassmannian and elliptic extensions of Dirac operators}

We have characterized the self--adjoint extensions of a given Dirac operator $D$ as the space
$\mathcal{M}$ of self--adjoint subspaces of a boundary Hilbert space $\mathcal{H}$ carrying
a polarization $\mathcal{H} = \mathcal{L}_-
\oplus \mathcal{L}_+$.   On the other hand, we have seen in the previous
section that the Grassmannian $\mathrm{Gr}'$ describes the
elliptic extensions of such operator.  Then, the elliptic self--adjoint
extensions of the given operators will be given by the intersection
$\mathcal{M} \cap\mathrm{Gr}'$.  This space will be called the
elliptic self--adjoint Grassmannian or the self--adjoint Grassmannian for
short.  It is possible to see that the
self--adjoint grassmannian is a smooth submanifold of the Grassmannian
and decomposes in connected components which are submanifolds of the
components $\mathrm{Gr}^{(k)}$.  We will denote the elliptic self--adjoint
grassmannian as $\mathcal{M}_{\ellip}$.  The most relevant topological and
geometrical aspects of $\mathcal{M}_{\ellip}$ are contained in the following
theorem.

\begin{theorem}  The elliptic self--adjoint Grassmannian is a Lagrangian
submanifold of the infinite dimensional Grassmannian.
\end{theorem}

{\parindent 0cm \emph{Proof:}}  That $\mathcal{M}_{\ellip}$ is an isotropic submanifold of 
$\mathrm{Gr} (\mathcal{L}_-, \mathcal{L}_+)$
follows immediately from Eq. \eqref{symp} and the observation that tangent vectors to
$\mathcal{M}_{\ellip}$ at
$W$ are defined by self--adjoint operators.  

Now, all we have to do is to compute $T_W \mathcal{M}_{\ellip}^\perp$ at $W = 0$
because of the homogeneity of the Grassmannian.  Hence, if $\dot A \in 
T_0 \mathcal{M}_{\ellip}^\perp$, this means that 
$$ \Tr (\dot A^\dagger \dot B - \dot B \dot A) = 0 ,$$  
for every self--adjoint $\dot B \in \mathcal{J}_2 (\mathcal{L}_-,Ê\mathcal{L}_+ )$, hence
$\dot A^\dagger - \dot A = 0$, and $\dot A$ is self--adjoint, then lying
in $T_0 \mathcal{M}_{\ellip}$. \hfill$\Box$

\bigskip
\bigskip

\paragraph{Acknowledgments}
This work was partially supported by MEC grant
MTM2010-21186-C02-02 and QUITEMAD programme. 


\end{document}